\documentclass[onefignum,onetabnum]{siamart190516}  
\usepackage{microtype}
\usepackage{amsfonts}
\usepackage{graphicx}
\usepackage{epstopdf}
\usepackage{amsmath, amssymb}
\usepackage{eufrak} 
\usepackage{yfonts}  
\usepackage[title]{appendix} 
\usepackage{mathtools}  
\usepackage{cleveref} 
\usepackage{enumerate} 
\usepackage{bm} 
\usepackage{soul}  
\usepackage[utf8]{inputenc} 
\usepackage[T1]{fontenc}
\usepackage[normalem]{ulem}
\usepackage{stmaryrd}
\usepackage{tikz}  
\usepackage[]{geometry} 
\numberwithin{equation}{section}
\usepackage{dsfont} 
\usepackage{graphicx}    
\usepackage{subcaption}

\ifpdf
  \DeclareGraphicsExtensions{.eps,.pdf,.png,.jpg}
\else
  \DeclareGraphicsExtensions{.eps}
\fi

\usepackage{enumitem}
\setlist[enumerate]{leftmargin=.5in} 
\setlist[itemize]{leftmargin=.5in}

\newsiamremark{remark}{Remark}
\newsiamremark{example}{Example}
\newsiamremark{hypothesis}{Hypothesis}
\crefname{hypothesis}{Hypothesis}{Hypotheses} 
\newsiamthm{claim}{Claim}
\newsiamthm{assumption}{Assumption}

\newcommand{\csol}{\widehat{\delta}(\vt^A, v^B)} 
\newcommand{\csolt}{\delta^*(p, X)} 
\newcommand{\btau}{\bar{\tau}} 

\newcommand{\gratio}{\frac{\gamma^B}{\gamma^A}}

\newcommand{\calY}{\mathcal{Y}}
\newcommand{\Vb}{\bar{V}}

\newcommand{\1}{\mathds{1}}

\newcommand{\dpdt}{\df \dsP \otimes \df t-\text{a.s}}
\newcommand{\calF}{\mathcal{F}}

\newcommand{\calA}{\mathcal{A}}
\newcommand{\calD}{\mathcal{D}}
\newcommand{\frD}{\textfrak{D}}

\newcommand{\frC}{\mathfrak{C}}

\newcommand{\calG}{\mathcal{G}}
\newcommand{\calP}{\mathcal{P}}
\newcommand{\calO}{\mathcal{O}}

\newcommand{\vt}{\tilde{v}}

\newcommand{\ft}{\tilde{f}}
\newcommand{\fh}{\hat{f}}

\newcommand{\dsE}{\mathbb{E}}
\newcommand{\dsF}{\mathbb{F}}
\newcommand{\dsH}{\mathbb{H}}
\newcommand{\dsQ}{\mathbb{Q}} 
\newcommand{\dsP}{\mathbb{P}}

\newcommand{\dsL}{\mathbb{L}}
\newcommand{\dsR}{\mathbb{R}}
\newcommand{\dsS}{\mathbb{S}}
\newcommand{\dsG}{\mathbb{G}}
\newcommand{\dsD}{\mathbb{D}}

\newcommand{\SUM}{\displaystyle\sum}

\newcommand{\IT}{^i_t}

\newcommand{\ph}{\hat{p}}

\newcommand{\agrc}{\textit{agreement-cost }}

\DeclareMathOperator*{\argmax}{arg\,max}
\DeclareMathOperator*{\esssup}{ess\,sup}
\usepackage{bbding}
\usepackage{thmtools}

\newcommand{\gt}{\tilde{g}}

\newcommand*\df{\mathop{}\!\mathrm{d}}

\newcommand{\Rom}[1]{\uppercase\expandafter{\romannumeral #1\relax}}
\newcommand{\rom}[1]{\lowercase\expandafter{\romannumeral #1\relax}}

\newcommand{\RN}[1]{%
  \textup{\uppercase\expandafter{\romannumeral#1}}%
}

\DeclareMathOperator{\diag}{diag} 
\DeclareMathOperator{\row}{row}     

\usepackage{amsopn}

\headers{A Risk-Sharing Framework of Bilateral Contracts}{J. Lee, S. Sturm, C. Zhou}

\title{A Risk-Sharing Framework of Bilateral
  Contracts\thanks{Junbeom Lee and Chao Zhou received the support
     from the French Ministry of Foreign Affairs and the Merlion programme as
     well as the Singapore MOE AcRF grants R-146-000-255-114 and R-146-000-243-114.}}

\author{Junbeom Lee\thanks{Department of Sales and Trading, Yuanta Securities Korea
    (\email{junbeoml22@gmail.com}). Most part of this work was carried out when
    the author was in National University of Singapore. Opinions in this paper
    are those of the author, and do not represent the view of Yuanta Securities Korea.}
  \and Stephan Sturm \thanks{Department of Mathematical Sciences,  Worcester
  Polytechnic Institute (\email{ssturm@wpi.edu}).}
  \and Chao Zhou \thanks{Department of
    Mathematics, National University of Singapore (\email{matzc@nus.ed.sg})}}

\ifpdf
\hypersetup{
  pdftitle={A Risk-Sharing Framework of Bilateral},
  pdfauthor={J. Lee, S. Sturm, C. Zhou}
}  
\fi

\begin{document}  

\maketitle

\begin{abstract}
  We introduce a two-agent problem which is inspired by price asymmetry arising
  from funding difference. When two parties have different funding rates, the
  two parties deduce different fair prices for derivative contracts even under
  the same pricing methodology and parameters. Thus, the two parties should
  enter the derivative contracts with a negotiated price, and we call the
  negotiation a \textit{risk-sharing} problem.  This framework defines the
  negotiation as a problem that maximizes the sum of utilities of the two
  parties. By the derived optimal price, we provide a theoretical analysis on
  how the price is determined between the two parties. As well as the price, the
  \textit{risk-sharing} framework produces an optimal amount of collateral. The
  derived optimal collateral can be used for contracts between financial firms
  and non-financial firms. However, inter-dealers markets are governed by
  regulations. As recommended in Basel \Rom{3}, it is a convention in
  inter-dealer contracts to pledge the full amount of a close-out price as
  collateral. In this case, using the optimal collateral, we interpret
  conditions for the full margin requirement to be indeed optimal.  
\end{abstract}



\section{Introduction}

In the aftermath of the financial crisis, it has become customary in recent
years for trading desks to make several adjustments in derivative transactions
for counterparty default risk, funding spreads, collateral, etc. For pricing
derivatives with the collective adjustments, many methodologies have been
developed by, e.g., \cite{piterbarg2010funding, burgard2010partial,
  brigo2011collateral, wu2015cva, crepey2017capital}. However, it is known that
the fair values derived by the developed methodologies are not fully recouped
from counterparties.  This can possibly due to inclusion of funding spread. For
traders, if the increased funding costs are not compensated from the
counterparty, it will be losses on the trades. However, considering the choices
of funding in derivative prices is a violation of Modigliani–Miller (MM)
theorem. For MM theorem to be valid, the absence of frictional financial
distress costs is required; see
\cite{modigliani1958cost,stiglitz1969re}. Therefore, considering funding
cost/benefit may be justified by market frictions.

Even so, there still remain some puzzles related to the funding adjustment.
First, when funding cost/benefit is accepted, it gives rise to asymmetry of
theoretical prices between two contractors even under the same pricing
methodology and parameters. The value fair to one party may not be fair to the
counterparty since funding rates of the counterparty is different from those of
the other party. Second, as asked by \cite{hull2012fva, hull2012fva2}, if
funding cost should be really considered, possibly due to market frictions, why
do banks buy Treasury bonds that return less than their funding costs?

Motivated by the related issues, we introduce a two-agent problem. Instead of
using the individual fair values, two parties may enter a contract through
negotiation by sharing costs, as briefly mentioned by \cite{li2016fva}. We
describe the negotiation problem as maximizing the sum of utilities of both
parties and we call this a \textit{risk-sharing} problem. Then, the
\textit{risk-sharing} framework theoretically analyzes how the price equilibrium
is determined and we can interpret the questions on funding adjustment by using
the derived price.  The optimal price from the
\textit{risk-sharing} framework will depend on the risk aversion parameters and
relative negotiation power between the two parties, but they are not observable
in markets. Therefore, the importance of this study is on providing sound
theoretical interpretations for the puzzles on funding difference.

The other part of the solution in the \textit{risk-sharing} framework is
collateral. In recent times, most OTC derivative contracts are
collateralized. There are multiple procedures for the margin, but in bilateral
contracts, it is general to post \textit{variation margin} and \textit{initial
  margin}. In our model, the focus is on \textit{variation margin} which traces
mark-to-market exposures. As stated in \cite[p.15]{bis2015margin}, ``for
variation margin, the full amount necessary to fully collateralise the
mark-to-market exposure of the non-centrally cleared derivatives must be
exchanged.'' This full collateralization on \textit{variation margin} has been 
settled as a market convention in inter-dealer transactions. On the other hand,
there is no such convention between banks and sovereign or corporate
clients. Indeed, it is partly or not collateralized for contracts between
financial firms and non-financial firms. Therefore, the \textit{risk-sharing}
framework provides the optimal amount of \textit{variation margin} for the
contract between a financial firm and non-financial firm. Since inter-dealer
contracts are governed by regulation in practice, we interpret the meaning
behind the margin requirement.

The optimal collateral in our model is represented by a certain stochastic
process. Thus, full \textit{variation margin} may not be optimal in
general. However, we do not conclude that the convention is unreasonable.
\textit{Variation margin} is posted on a daily or intra-day basis. If the amount
was calculated by a complicated rule at each time, the amount would be
unacceptable for some parties and this can be a possible cause of
conflict. Hence, rather than coming to a sensitive conclusion, we analyze the
situation for the margin requirement to be optimal. The market convention will
turn out to be based on certain conditions on funding cost/benefit considered in
derivative prices and hedging strategy taken by two parties. Especially, we will
see later that the full margin requirement is related to the absence of market
friction. 
   
One mathematical difficulty to deal with the \textit{risk-sharing} problem is
that the amount by breach of contract is given by piece-wise concave
functions. Mathematically similar problems were solved by
\cite{carassus2009portfolio, bo2016optimal, bo2017portfolio,
  yang2017constrained}.  In \cite{carassus2009portfolio}, portfolio optimization
problems were considered where the agent switches utilities. They used duality
method that cannot be applied to our problem as we cannot impose a positive
constraint for the portfolio. In \cite{bo2016optimal, bo2017portfolio}, the
piece-wise concave property arose from different lending/borrowing rates and
they solved the optimization problem by using HJB equations. In their problem,
the associated HJB equations had a homothetic property. Moreover, with a mild
assumption that the lending rate is smaller than the borrowing rate, the
Hamiltonian became continuous in their cases. However, in our problem, we deal
with two state processes taken by two utilities, so we cannot make use of a
similar approach.

We circumvent the above difficulties by imposing some conditions on funding
spread depending on choices of utilities.  For the funding spread, we assume
that the lending and borrowing rates are the same for each party.  To be more
precise, the two parties fund themselves on their own funding rate which may not
be the same as OIS rate, but the lending and borrowing rates are the
same. Moreover, the funding costs/benefits for delivering collateral of one or
both parties will be ignored for characterization of the optimal solution. More
precisely, we will examine two cases. First, we will consider two risk-averse
agents whose funding rates for delivering margin are OIS rate. Second, we will
also consider one risk-averse agent and one risk-neutral agent, and in this
case, the funding rate of the risk-neutral agent does not need to be OIS
rate. To streamline this paper, we mainly deal with the two risk-averse agents
in main sections and report the second case in \cref{app:sec:risk.neutral}. 

This funding condition can be understood that the party is an entity which
invests the capital without or with a small leverage, or the party can post
collateral with secured funding. Even though the secured funding for
\textit{variation margin} is not so general, some realistic cases are discussed
by \cite{albanese2013restructuring}. In addition, This setup on funding spread
can be partly justified by the results in \cite{lee2017binary} which showed
that, in many classes of derivatives, hedgers do not need to switch funding
state between lending and borrowing positions. In particular, it is guaranteed
that if a hedger does not enter borrowing state and the lending rate is same as
OIS rate, we can ignore the funding impacts.

In our model, we include default risk, funding
spread, and collateral. We consider incomplete markets that
hedgers cannot access to assets for hedging default risk such as bonds and
CDSs. The reference filtration is generated by a Brownian motion. The
mark-to-market exposure is calculated as \textit{clean price} which is the
classical risk-neutral price without default risks and funding spread. Moreover,
for risk-averse agents, we consider exponential utilities.
For simplicity, we consider non-incremental cash-flow in main sections and the 
incremental effect is discussed together with the case of a risk-neutral agent
only in \cref{app:sec:risk.neutral}. Then, this paper is organized as follows.  

In \cref{sec:modelling}, the \textit{risk-sharing} problem is introduced.  We
start from defining a filtration and making an assumption on default intensities
in \cref{sec:mathematical-setup}. Before giving the details, we
explain our motivation with a simple model in \cref{sec:motivation}.  Then we
describe cash-flow which are determined by dividends, margins, and close-out
amount. Both parties entering the contract will have a portfolio given in
\cref{sec:portfolio} depending on the cash-flow. The introduced 
\textit{risk-sharing} problem is maximizing the sum of utilities of discounted
portfolio values at termination of the contract. In \cref{sec:reduction}, the
original form of the \textit{risk-sharing} problem is reduced so that it is
represented on the reference filtration. Then we define admissible sets more
precisely with this reduced problem.  We mainly deal with the reduced problem in
this paper. In \cref{sec:optimal.collateral}, we define a dynamic version of the
main problem, and  optimal collateral is characterized. Then given the
optimal collateral, we derive a condition to find an optimal price in
\cref{sec:optimal.price} and  examples are 
given in \cref{sec:example}.
\section{Modeling}
\label{sec:modelling}  
\subsection{Mathematical Setup}
\label{sec:mathematical-setup}
We consider two parties entering a bilaterally cleared contract. We call the two
parties ``Agent A'' and ``Agent B'', respectively. In what follows, an index $A$
(resp. $B$) is used to stand for the Agent A (resp. Agent B). We consider a
probability space $(\Omega, \calG, \dsP)$ with physical probability $\dsP$ and let
$\dsE$ be the expectation under $\dsP$. For $i \in \{A, B\}$, we define
non-negative random variables $\tau^i$ on $(\Omega, \calG, \dsP)$ such that
$\dsP(\tau^i = 0) = 0$ and $\dsP(\tau^i > t) >0$, for any  $t\geq0$, to represent default
times of the agents. We let
\begin{align}
\tau \coloneqq \tau^A \wedge \tau^B, ~~~~~ \btau \coloneqq \tau \wedge T, \nonumber
\end{align}
where $T>0$ is the maturity of a certain derivative contract.  We denote by
$(W_t)_{t\geq0}$  a $d$-dimensional standard Brownian motion under
$\dsP$. The reference filtration $\dsF = (\calF_t)_{t \geq0}$ is the \textit{usual
  natural filtration} of $(W_t)_{t \geq0 }$, and the full filtration $\dsG$ is
defined as
\begin{align}
  \dsG = (\calG_t)_{t \geq0}
  = \Big(\calF_t \vee \sigma\big(\{\tau^i \leq u\}: u \leq t, i \in \{A, B\}\big)\Big)_{t
  \geq0}.\nonumber 
\end{align}
Then, we consider a filtered probability space $(\Omega, \calG, \dsG, \dsP)$. Note
that for any $i \in \{A, B\}$, $\tau^i$ is a $\dsG$-stopping time but may fail to be
an $\dsF$-stopping time.  Unless stated, every process is a
$(\dsP, \dsG)$-semimartingale.
%
As a convention, for any $\dsG$-progressively measurable process
$(u_t)_{t \geq0}$ and $(\dsP, \dsG)$-semimartingale $(U_t)_{t\geq0}$,
 $\int_{s}^{t}u_s\df U_s = \int_{(s,t]}^{}u_s\df U_s$,
where the integral is well defined. In addition, for any $\dsG$-stopping time $\theta$ and process $(\xi_t)_{t \geq0}$,
  we denote  
  \begin{align}
     \xi^\theta_\cdot \coloneqq \xi_{\cdot \wedge \theta},     \nonumber
  \end{align}
  and when $\xi_{\theta-}$ exists, denote
  $\Delta \xi_\theta\coloneqq \xi_\theta - \xi_{\theta-}$. For $i \in \{A, B\}$,
  $t \geq0$, we also let
  \begin{align}
  G\IT \coloneqq \dsP(\tau^i > t|\calF_t) ~~\text{and}~~
  G_t\coloneqq  \dsP(\tau > t|\calF_t).  \nonumber
  \end{align}
The following assumption stands throughout this paper.
\begin{assumption}\label{assm:intensity}
\begin{enumerate}[label=(\roman*)]
\item $(G_t)_{t \geq0}$ is non-increasing and absolutely continuous
  with respect to Lebesgue measure. 
\item For any $i \in \{A, B\}$, there exists a process $h^i$, defined as
\begin{align}
  h^i_t \coloneqq \lim_{u\downarrow0}\frac{1}{u}\frac{\dsP(t < \tau^i \leq t+u, \tau > t|\calF_t)}{\dsP(\tau >
  t|\calF_t)}, \nonumber 
\end{align}
and
$ (M^i_t)_{t\geq0} \coloneqq \big(\1_{\tau^i \leq t \wedge \tau} - \int_{0}^{t \wedge \tau}h^i_s \df s\big)_{t\geq0}$
is a $(\dsP, \dsG)$-martingale.
\end{enumerate}
\end{assumption}  
We denote $h \coloneqq h^A + h^B$. By (\rom{1}) in \cref{assm:intensity}, there
exists an $\dsF$-progressively measurable process $(h^0_t)_{t\geq0}$ such that
\begin{align}
   h^0_t = \lim_{u\downarrow0}\frac{1}{u}\frac{\dsP(t < \tau \leq t+u|\calF_t)}{\dsP(\tau >
  t|\calF_t)},  \nonumber 
\end{align}
and 
$(M_t)_{t\geq0}\coloneqq\big(\1_{\tau\leq t} - \int_{0}^{t\wedge\tau}h^0_s\df s\big)_{t\geq0}$ 
is also a $(\dsP, \dsG)$-martingale.  When $\tau^A$ and $\tau^B$ are independent on
$\dsF$
\begin{align}
h^\Delta\coloneqq h - h^0=0. \nonumber  
\end{align}
In general, it is not the case.  Moreover, by (\rom{1}) in \cref{assm:intensity}
and \cite[Corollary 3.4]{coculescu2012hazard}), $\tau$ avoids any
$\dsF$-stopping time. In other words, for any $\dsF$-stopping time $\tau^\dsF$,
\begin{align} 
 \dsP(\tau = \tau^\dsF) = 0.  \label{avoid}
\end{align}
\begin{remark}
  It is worth discussing the meaning of the item (\rom{1}) in
  \cref{assm:intensity} in view of both modeling and mathematical
  aspects. Without the assumption, $G$ is only $(\dsP, \dsG)$-supermartingale,
  thus, by the Doob-Meyer decomposition, there exist a $(\dsP, \dsG)$-martingale
  $\nu$ and non-increasing process $\upsilon$ such that $G = \nu + \upsilon$. Therefore, assuming
  $G$ is non-increasing is equivalent to setting $\nu =0$, i.e., we ignore some
  parts of random effects in default times for mathematical simplicity.  In
  addition, the condition (\rom{1}) is equivalent to the statement that for any
  $\dsF$-martingale $\xi$, the stopped process $\xi^\tau$ is a
  $\dsG$-martingale, see Proposition 3.4 in \cite{elliott2000models}. Thus, the
  condition is close to ($H$)-hypothesis. 
\end{remark}
On this setup, we can reduce the full filtration using \cref{lem:red} reported
in \cref{sec:lemma}. For denoting the spaces of random variables and processes,
we use standard notations which are given in \cref{sec:spaces}.
\subsection{A Motivation}
\label{sec:motivation}
Before delving into the details, we will explain a motivation of our
\textit{risk-sharing} problem with a simple model. Let us consider two agents,
$A$ and $B$ with constant funding rates $R^A$ and $R^B$. We moreover, consider a
situation that the Agent A buys from the Agent B an uncollateralized bond of
unit notional amount and maturity $T$. If the two agents were able to fund by
the (so-called) risk-free rate $r$, the fair value of the two parties would be
$e^{-rT}$.  However, when $R^i$, $i \in \{A, B\}$, are not equal to $r$, the fair
values of two parties are different to the parties, and the two parties would
want to recoup their individual adjustments: $(e^{-R^iT} - e^{-rt})$.  Given the
asymmetry by funding difference, we want to model how the price is
determined.

To this end, let $p$ be the adjustment price ``given to $A$'' on top of the (clean)
risk-neutral price $e^{-rT}$, e.g., for the bond contract, $A$ pays $-e^{-rT} +
p$ to $B$ at initiation of the contract. This money is invested in their funding
accounts up to $T$, and at the maturity, the Agent A will receive $1$ dollar
amount from the Agent B. Therefore, the respective profit and loss of the two parties at
$T$ will be
\begin{align}
  V_T^{A, p} = (-e^{-rT} +p)e^{R^AT} + 1, ~~~ V_T^{B, p} = (e^{-rT} -p)e^{R^BT} - 1.
  \nonumber  
\end{align}
For the time being, we assume that the two parties both have the same preference of
exponential utility as $U(x) = -e^{-x}$. Then, we find an optimal adjustment
price $p^*$ to maximize the two parties' aggregated utility of 
discounted P/L by their own funding rates, namely, 
\begin{align}
 p^* = \argmax_{p\in \dsR} \bigg[U\big(-e^{-rT}+p+e^{-R^AT}\big)
  + \lambda U\big(e^{-rT}-p-e^{-R^BT}\big)   \bigg], \label{first.modeling}
\end{align}
for some $\lambda>0$. The parameter $\lambda$  can be interpreted as a relative bargaining
power of $B$. By straightforward calculation, \cref{first.modeling} becomes
\begin{align}
  p^* = -\frac{e^{-R^AT} + e^{-R^BT}}{2} + e^{-rT} - \frac{\ln{(\lambda)}}{2}.
    \label{firstex}
\end{align}
From \cref{firstex}, if the two agents have the same
negotiation power, i.e., $\lambda=1$, the optimal adjustment price $p^*$ is determined
as the middle of the individual adjustments, i.e.,
\begin{align}
p^*=  -\frac{1}{2}\Big[(e^{-R^AT} - e^{-rT})   +(e^{-R^BT} - e^{-rT})\Big]. \label{mid.price}
\end{align}
In addition, as $\lambda$ increases, the contract becomes more advantageous to the
Agent B. In the case that the Agent B is a government, $\lambda$ can be large possibly
due to tax benefits in buying treasury securities. However, it should be
mentioned that $\lambda$ is generally not observable in markets, so the importance of
our model mainly remains in  theoretical analysis. In the following sections, we
describe the agents' P/L with more details in terms of hedging portfolios for
entering a derivative contract.
\begin{remark}
  A condition of \textit{funding transfer policy} (FTP) that is beneficial to
  both parties was also discussed by \cite{albanese2018wealth}. It was shown
  that (\ref{mid.price}) is one of the choices satisfying their condition; see
  \cite[Proposition 5.1]{albanese2018wealth}. However, there may be many choices
  of the FTP satisfying the condition, so instead, we investigate the prices
  which are the best to the parties.
\end{remark}  
\subsection{Hedging Portfolio under Bilateral Contracts}
\label{sec:portfolio}
In this section, under CVA, DVA, funding spread, and collateral, we define the
two parties' hedging portfolios for entering a contract. We mostly depict the
hedging portfolio in view of $A$. Then the portfolio of $B$ can be derived by
a similar way.  
\subsubsection{Dividend, Close-out Amount, and Collateral}
We begin this section by explaining the cash-flow in bilateral
contracts. Consider two agents who want to enter a bilateral contract which
exchanges promised dividends. We denote the cumulative dividend process by
$\frD$. We assume that $\frD$ is an $\dsF$-adapted c\`adl\`ag process and
is independent of defaults. The value of $\frD$ is determined by an
$n$-dimensional $\dsF$-adapted (i.e., non-defaultable) underlying asset
$S = (S^1, \dots, S^n)$ that satisfies the following stochastic differential
equation (SDE):
\begin{align}
 \df S\IT = \mu^i_tS\IT \df t + (\sigma\IT)^\top S\IT \df W_t,~~1 \leq  i \leq n,  \label{irep}
\end{align}
where $\sigma^i \in \dsR^{d}$ and $\mu^i \in \dsR$ are $\dsF$-predictable. Moreover, we denote $\mu \in
\dsR^n$ and $\sigma\in \dsR^{n \times d}$ such that $(\mu)_i = \mu^i$ and $\row(\sigma)_i = \sigma^i$.

It is not assumed that $n = d$. In other words, the considered market may or may
not be complete regardless of whether assets to hedge default risk, such as CDSs
and bonds, are traded. In this paper, we consider markets with the absence of
assets to hedge the default risk. We only assume that for all $t$, $\sigma_t$ is of
full rank so that we can define the risk premium $\Lambda$ as a solution of
\begin{align}
 \sigma\Lambda =   (\mu - r\1 ), \label{fullrank}
\end{align}
where $\1 \coloneqq (1, \dots, 1) \in \dsR^d$ and $r$ is an $\dsF$-adapted process
which represents overnight indexed swap (OIS) rate. We will later use $\Lambda$ for a
pricing measure to define close-out amount. Recall that the existence of
$\Lambda$ guarantees \textit{arbitrage-free} condition in classical context. However,
since the classical definition of \textit{arbitrage opportunity} does not
reflect adequately the hedger-specific nature of bilateral contracts, there have
been many studies to redefine \textit{arbitrage opportunity} properly in the
context of bilateral contracts. The condition being developed is slightly
different from paper to paper, but often absence of \textit{arbitrage
  opportunity} is obtained with similar conditions to (\ref{fullrank}). See, for
example, \cite[Proposition 3.3]{bielecki2015valuation}. For definitions of 
hedger-specific \textit{arbitrage opportunities}, readers can refer to
\cite{bielecki2015valuation, bielecki2018arbitrage, bichuch2016arbitrage1,
  bichuch2016arbitrage2, bichuch2017arbitrage, nie2018game, nie2018american}. 

We set, as a convention, a positive value (resp. negative) of dividend process at
a certain moment to mean that the Agent A pays to (resp. is paid by) the
Agent B. For example, if $A$ sells a put option on $S$ with the 
exercise price $\kappa$ and maturity $T$, then for any $t \geq0$,
 $\frD_t = \1_{t \geq T}(\kappa-S_T)^+$. 
Note that the initial price exchanged at  initiation of the contract is not a
part of $\frD$. We will include the initial price in the hedging portfolios as
their initial value.  Because jumps of an $\dsF$-adapted c\`adl\`ag process
are exhausted by $\dsF$-stopping times by \cite[Theorem
4.21]{he1992semimartingale}, and $\tau$ avoids any $\dsF$-stopping time (recall
(\ref{avoid})), $\frD$ does not jump at default, i.e.,
\begin{align}
\Delta\frD_{\tau} = 0,\quad \text{almost surely.} \label{djump}
\end{align}

Let us turn to explain close-out amount and margin process. The obligation on
dividend $\frD$ may not be fully honored at one party's default. For the default
risk, covenants of the close-out amount and collateral are documented in a
Credit Support Annex\footnote{A part of ISDA Master Agreement.}  before
initiation of the contract. At the event of default, the dividend stream stops
and CSA close-out amount should be settled. However, because of the default, the
defaulting party would not be able to pay the full close-out amount. To mitigate
the risk of losses at default, collateral is exchanged between the two
parties. In bilateral contracts, \textit{variation margin} and \textit{initial
  margin} are posted in general, and the close-out amount is often determined as
mark-to-market exposure. 

In our model, only \textit{variation margin} is a part of the control variables
of our stochastic control problem that will be introduced later.  We exclude
\textit{initial margin} for simplicity.   As stated  
in \cite[p.12]{bis2015margin}, ``the amount of \textit{variation margin}
reflects the size of this current exposure,'' and it is recommended that ``the
full amount necessary to fully collateralise the mark-to-market exposure of the
non-centrally cleared derivatives must be exchanged''
\cite[p.15]{bis2015margin}. The meaning behind this regulation will be discussed   
later based on our model.

\begin{remark}
For general practices of \textit{initial margin}, readers can refer to
\cite{murphy2013otc, brigo2014ccp}. In addition, \textit{initial margin} causes
associated BSDEs anticipated. For the numerical simulation of anticipated BSDEs,
readers can refer to \cite{agarwal2019numerical}. 
\end{remark}

Now, we depict the close-out amount and \textit{variation margin}
mathematically. One of the popular choices to calculate the market exposure is
\textit{clean price} which is basically the classical risk-neutral price. As
used in the classical pricing, we use the ``so-called'' risk-free rate.  Note
that it is a little out of context to call it the risk-free rate since arguments
under bilateral contracts are from the reality that dealers cannot access to
the risk-free rate, yet it is acknowledged that OIS rate is the best proxy for
the so-called risk-free rate. Thus, in what follows, we use OIS rate for
evaluating the \textit{clean price}, and denote it by $(r_t)_{t\geq0}$. We assume $(r_t)_{t\geq0}$ is
$\dsF$-adapted and denote by $B$  the  money market account on $(r_t)_{t\geq0}$, namely
\begin{align}
 B_t \coloneqq \exp{\bigg(\int_{0}^{t}r_s\df s\bigg)} ~~\text{for any } t \geq0.\nonumber 
\end{align} 
We can find the pricing measure $\dsQ$ such that $B^{-1}S^i$, $1 \leq i \leq n$, are
$(\dsQ, \dsF)$-local martingales because $\sigma$ is of full rank as in
(\ref{fullrank}).  Let $e_t$ denote the new mark-to-market exposure at
$t \leq T$. We assume that the market exposure is calculated as \textit{clean
  price}:
\begin{assumption}\label{assm:cp}
For any $t\in [0, T]$, $e_t = B_t\dsE^{\dsQ}\Big(\int_{t}^{T}B^{-1}_s\df \textfrak{D}_s\Big|\calF_t\Big)$.
\end{assumption}

By \cref{assm:cp}, we can derive some properties of $(e_t)_{t\geq0}$.  We report
the proof in \cref{app:sec:proof}.
\begin{lemma}\label{lem:em}
\begin{enumerate}[label=(\roman*)]   
\item $e _T = 0$.
\item $e _{\btau} = \1_{\tau \leq T}e _\tau$.
\item   $\df e _t = \big(r_te _t +  B_tZ _t\Lambda_t\big)\df t +B_tZ_t\df W_t - \df
  \frD _t$, for $ t \in [0, T]$,  for some $Z  \in \dsH^{2, d}_{T, loc}$. 
\item $e _{\tau-} = e _{\tau}$ almost surely.
\end{enumerate}
\end{lemma}
\begin{remark}
Notice that $Z $ is closely related with the delta risk of
$\tilde{e} _t = B^{-1}_te _t$. Indeed, if $e $ is Malliavin differentiable
and is a smooth function of $S_t$, i.e., $\tilde{e} _t = \tilde{e} (t, S_t)$, then by Clark-Ocone formula,
 $Z _t = D_t\tilde{e} _t = [\diag (S_t)\sigma_t]^\top \nabla_S \tilde{e} (t, S_t)$,
where $D_t$ is the Malliavin derivative operator at $t$ and $\nabla_S$ is the
classical gradient. We shall see later that $Z $ has a special role to interpret the meaning behind
full collateralization. In addition, if $\frD  \in \dsS^2_T$, we can choose $Z $
in $\dsH^{2, d}_T$.
\end{remark}

Let $m _t$ denote the amount of \textit{variation margin} posted at $t \leq
T$. Similarly, $m _t \geq0$ (resp. $m _t< 0$) means that $A$ posts
(resp. receives) the margin to $B$ at time $t \leq T$.  We assume $(m _t)_{t\geq0}$ is 
an $\dsF$-adapted c\`adl\`ag process.  Note that $m $ is chosen to be
$\dsF$-adapted for consistency in financial modeling. The collateral is required
because we do not know the full information of default. Therefore, the amount of
collateral is calculated only by available information $\dsF$.  The
admissible set will be defined more precisely when the \textit{risk-sharing}
problem is introduced.

Once one party announces bankruptcy, the margin process stops. Therefore, at the
default $\tau \leq T$, the amount of collateral is $m _{\tau-}$, but wealth which amounts
to $e _{\tau}+\Delta \frD _{\tau} ~(= e _{\tau} \text{ a.s})$ should be transferred from
$A$ to $B$. In addition, the loss by breach of the contract will be inflicted to
the Agent B (resp. A) only when $\tau = \tau^A$ and
$e _\tau \geq m _{\tau-}$ (resp. $\tau = \tau^B$ and $e _{\tau } < m _{\tau-}$). Denoting
the loss rate of the Agent A (resp. B) by $L^A$ (resp. $L^B)$, the amount
by breach is
\begin{align}
\1_{\tau =\tau^A}L^A(e _{\tau} - m _{\tau-})^+ - \1_{\tau=\tau^B}L^B(e _{\tau} - m _{\tau-})^-. \nonumber 
\end{align}
We assume that $L^i$, $i \in \{A, B\}$, are positive constant. Finally, we can define the
full cash-flow $\frC $ as
\begin{align}
  \frC _t \coloneqq
  & \1_{\tau > t}\frD _t + \1_{\tau \leq t}\big(\frD _{\tau} + e _{\tau}\big)\nonumber\\
  &-\1_{\tau =\tau^A\leq t}L^A(e _{\tau} +\Delta\frD _{\tau}- m _{\tau-})^+ + \1_{\tau=\tau^B\leq t}L^B(e _{\tau}
  +\Delta\frD _{\tau} - m _{\tau-})^-.\nonumber 
\end{align}
By (\ref{djump}) and the last item in \cref{lem:em}, for any $t \leq T$, almost surely
\begin{align}
  \frC _t =&\1_{\tau > t}\frD _t + \1_{\tau \leq t}\big(\frD _{\tau} + e _{\tau}\big)\nonumber\\
  &-\1_{\tau =\tau^A\leq t}L^A(e _{\tau} - m _{\tau-})^+ + \1_{\tau=\tau^B\leq t}L^B(e _{\tau} - m _{\tau-})^-.\label{cashflow}
\end{align}

In the next section, we define a self-financing portfolios to hedge against $\frC$
with more details. We construct the portfolio in view of the Agent A since the portfolio
of the Agent B is in most ways similar. Before proceeding, we
provide some remarks related to possible extensions of our model.
\begin{remark}
\begin{enumerate}[label=(\roman*)]
\item One may argue that \textit{clean
    price} is not an appropriate close-out amount since the Agent B's
  default is not considered.  However, taking default risk of the Agent B
  into the exposure may heavily penalize the surviving party, because the
  default event of one party can  negatively affect the creditworthiness of the
  surviving party, especially when the defaulting member has an impact on
  systemic risk. For such discussion, readers can refer to
  \cite{brigo2011close}.
\item In practice, \textit{variation margin} is called on intra-day basis (say,
  two or three times per day). In this paper, we assume a continuous margin
  process for simplicity. One may want to model \textit{variation margin} as a
  c\`adl\`ag step process to describe reality more precisely, cf. \cite{brigo2014nonlinear}.
\item Underlying assets subject to defaults are beyond the scope of this
  paper. For modeling with emphasis on contagion risk, readers may want to refer
  to \cite{jiao2013optimal, bo2019locally, brigo2014arbitrage, bo2017credit}.      
\item In reality, it is hard to estimate exact default intensities. For example,
  dependence between the Agent B's exposure and default probability is not
  negligible, which is sometimes called right/wrong way risk, but it is
  challenging to estimate the dependence from market quotation. Thus, such
  issues lead to robust pricing arguments. See \cite{glasserman2018bounding,
    bichuch2018robust}.
\end{enumerate}   
\end{remark}

\subsubsection{Self-Financing Hedging Portfolio}
\label{sec:hedging.portfolio}
\begin{figure}
  \centering
\begin{tikzpicture}
    \draw [blue,thick](-1.8,0) ellipse (1.2 and 0.7);
    \node [] at (-1.8,0) {$A$};   
    \draw [blue,thick](1.8,0) ellipse (1.2 and 0.7);
    \node [] at (1.8,0) {$B$};
    \draw [dashed, thick, <->] (-0.5, 0) -- (0.5,0);
    \node [below] at (0,0) {$r^m$};
    \draw [dashed, thick, <->] (-4., 1.2) -- (-3., 0.4);
    \node [right] at (-3.5, 1.1) {$R^{A, m}$};
    \draw [dashed, thick, <->] (4., 1.2) -- (3., 0.4);
    \node [left] at (3.6, 1.) {$R^{B, m}$};
    \draw [thick, black] (-7.2, 1.2) rectangle ++(3, 1);
    \node [] at (-5.7, 1.7) {Margin provider};
    \draw [thick, black] (-7.2, -2.2) rectangle ++(3, 1);
    \node [] at (-5.7, -1.7) {Funding desk};
    \draw [thick, black] (4.2, 1.2) rectangle ++(3, 1);
    \draw [thick, black] (4.2, -2.2) rectangle ++(3, 1);
    \node [] at (5.7, 1.7) {Margin provider};
    \node [] at (5.7, -1.7) {Funding desk};
    \draw [dashed, thick, <->] (-4., -1.2) -- (-3., -0.4);
    \node [right] at (-3.6, -1.) {$R^{A}$};
    \draw [dashed, thick, <->] (4., -1.2) -- (3., -0.4);
    \node [left] at (3.6, -1.) {$R^{B}$}; 
\end{tikzpicture}
\caption{Interest rates among parties. When a party is funded from its internal
  funding provider for delivering collateral, $R^{i, m} = R^i$.  In practice,
  $r^m$ is often chosen as federal funds rate or EONIA rate, i.e., $r^m = r$.}
\label{fig.rates}
\end{figure}
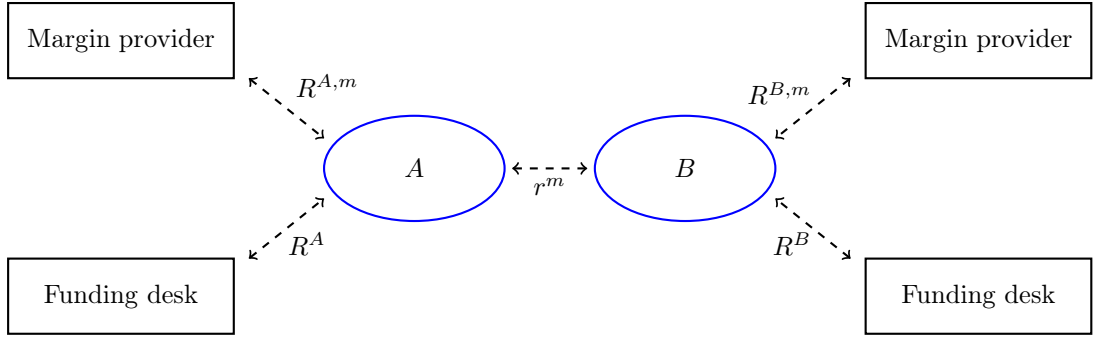
The funding sources of an agent can be an external funding provider, treasury 
department, repo markets, etc. After the financial crisis, such funding rates do
not represent the risk-free rate (approximately OIS rate in recent times). We
consider $\dsF$-adapted processes $(R^{i, m}_t)_{t\geq0}$, $ i \in \{A, B\}$, to
represent the margin funding rates offered from margin lenders, and denote that
for $t\geq0$,
\begin{align}
  B^{i, m}_t \coloneqq& \exp \Big(\int_{0}^{t}R^{i, m}_s \df s\Big). \label{mda}
\end{align}
Another cash-flow stream associated with margin process is remuneration from
margin receivers. When \textit{variation margin} is pledged by (resp. received by)
the Agent A, the Agent B should remunerate (resp. should be remunerated by)
the  Agent A with respect to an interest rate. We let $r^m$ and $B^m$ denote the
remuneration rate and account of the  Agent A. Therefore, for each party, the net
cost/benefit involved in posting the margin is determined by $R^{i, m} - r^m$,
$i \in \{A, B\}$, and we denote this spread by $s^{i, m}$, i.e.,
\begin{align}
  s^{i, m} \coloneqq R^{i, m} - r^m, ~~i \in \{A, B\}. \label{dms}  
\end{align}
In general, \textit{rehypothecation} is allowed
for \textit{variation margin}, in other words, the margin account can be used to
maintain the hedging portfolio.
We moreover, assume that the two parties should finance their operations by interest rates $R^i$,
$i\in \{A, B\}$, for constructing the rest of the portfolios. We
denote the associated funding account and spread by $B^i$ and $s^i$, $i \in \{A,
C\}$, respectively,
i.e.,
\begin{align}
 B^{i}_t \coloneqq& \exp \Big(\int_{0}^{t}R^i_s \df s\Big), \label{fa}\\
  s^{i}_t \coloneqq& R^{i}_t - r_t. \label{def.fs}
\end{align}

Then, each party constructs their hedging portfolio using the above accounts and
risky assets $(B^i, B^{i, m}, B^m, S)$, $i \in \{A, B\}$. Let
$\varphi^i\coloneqq (\eta^i$, $\eta^{i, m}$, $\eta^m$, $\eta^{i, S})$ denote the respective units
of $(B^i, B^{i, m}, B^m, S)$, $i \in \{A, B\}$, in the hedging portfolios and we
call $\varphi^i$ the trading strategy of the Agent $i$.  We assume that
$\varphi^i$, $i \in \{A, B\}$, are $\dsG$-predictable and use the convention that a
positive unit of trading strategy means long position.

If the Agent A posts collateral which amounts to $m_t$ at $t$, she needs to
deliver $\eta^{A, m}_t$ shares of the account $B^{A, m}_t$ from the margin
lender. Then, the Agent A will have $\eta^m_t$ shares of the margin account
$B^m_t$ to which the remuneration from the Agent B is accrued. Thus, 
\begin{eqnarray} 
  &\eta^mB^m = m , \label{margin.e1}\\
  &\eta^{A, m}B^{A, m} + \eta^mB^m = 0, \label{margin.e2}\\
  &\eta^{B, m}B^{B, m} - \eta^mB^m = 0. \label{margin.e3} 
\end{eqnarray}
\begin{remark}
As in practice, we consider cash collateral which is
\textit{rehypothecated}. Sometimes it is possible that risky assets can be
posted as collateral and the margin account is \textit{segregated}, which
means that the account is not included in the hedging portfolio. When we
consider a different convention, the mathematical structure of wealth process
also become different. For the various conventions, readers may want to refer
to \cite{bielecki2015valuation, crepey2014counterparty}. In addition,     
the amount of collateral may depend on the value of the hedging
portfolio, which is sometimes called endogenous
collateralization. \cite{burgard2010partial} and \cite{nie2016bsde} discussed
endogenous collateralization by PDEs and BSDEs, respectively.      
\end{remark}
Now, we are in a position to define a \textit{self-financing portfolio}.
\begin{definition}
If $V^A_t=V^A_t(\varphi^A, \frC)$, $t\in [0, T]$, defined as
\begin{align}
   V^A_t = \eta^{A}_tB^A_t + \eta^{A, m}_tB^{A, m}_t+\eta^{m}_tB^{m}_t+\eta^{A, S}_t  S_t, \label{self}
\end{align}
satisfies
\begin{align}
  V^A_t =&V^A_0 +\int_{0}^{t \wedge \btau}\eta^{A}_s\df B^A_s + \int_{0}^{t\wedge \btau}\eta^{A,
  m}_s\df B^{A, m}_s+ \int_{0}^{t\wedge \btau}\eta^{  m}_s\df B^{m}_s\nonumber\\
  &+\int_{0}^{t\wedge \btau}\eta^{A, S}_s \df S_s - \frC_{t\wedge \btau}, \label{selfdiff}
\end{align}
for any $t \in [0, T]$, then $V^A$ is called the \textit{self-financing portfolio}
of the Agent A.  
\end{definition}
\begin{remark}
 Note that for $ t > \btau$, $V_t^A=V^A_{\btau} $. In \cref{selfdiff}, $\frC_{t
 \wedge \btau} = \frC_t$, for any $ t\geq0$, by the definition \cref{cashflow}. 
\end{remark}
The \textit{self-financing portfolio} of the Agent B is defined similarly. The
difference is the direction of \textit{variation margin} and $\frC$. Then, by
 (\ref{dms}), (\ref{margin.e1})-(\ref{selfdiff}), we can see that 
\textit{self-financing} portfolio processes of the  Agent A and, similarly,
the Agent B follow
\begin{align}
  \df V^A_t =& \Big(R^A_tV^A_t-s^{A, m}_tm _t + \eta^{A, S}_t\odot S_t[\mu_t - \1 R^A_t]\Big)\df t +
               \eta^{A, S}_t\odot S_t \sigma_t\df W_t -
               \df \frC_t, \label{hph} \\
  \df V^B_t =& \Big(R^B_tV^B_t~+s^{B, m}m _t +  \eta^{B, S}_t\odot S_t[\mu_t - \1
               R^B_t]\Big)\df t
               ~+ \eta^{B, S}_t\odot S_t\sigma_t \df W_t+\df \frC_t, \label{hpc}
\end{align}
where $\odot$ is component-wise product. If we consider an agent who has a naked
position against market risk, we set $\eta^{i, S} = 0$.  Before examining whether
(\ref{hph}) and (\ref{hpc}) are well defined, we first want to introduce our
target problem.

We find the best initial price and amount of \textit{variation margin} to
optimize the aggregated utilities of both parties. If there were no adjustment
in pricing, the classical risk-neutral price $e _0$ should be exchanged at
initiation of the contract. Let $p$ denote the amount paid to the Agent A on top
of $e _0$. Therefore, initial price paid to the Agent A is $e _0 + p$. More
precisely, denoting the \textit{initial endowment} of each party by $\nu^A$ and
$\nu^B$,
\begin{align}
V^A_0 =  \nu^A + e _0 + p, ~~\text{and}~~V^B_0 = & \nu^B - e _0 - p. \nonumber
\end{align}
Thus, $V^i$ depends on the choice of $(p, m )$. For simplicity of notations, we
often suppress $(p, m )$, e.g., $V^i = V^{i, p, m },~i \in \{A, B\}$.  Then, with
an admissible set $\calA$, utilities $U_i\colon \dsR \to \dsR$, and $\lambda>0$, we
define the \textit{risk-sharing} problem as follows:
\begin{align}
  (p^*, m^*) =\argmax_{(p, m ) \in \calA}\dsE\Big[U_A\big((B^A_{\btau})^{-1}V^{A, p, m }_{\btau}\big)
  + \lambda U_B\big((B^B_{\btau})^{-1}V^{B, p, m }_{\btau}\big) \Big]. \label{firstdef}
\end{align}
We will define $\calA$ more precisely in the following section.  Note that hedging
strategies are not control variables. In other words, we assume that two parties
choose their strategies by their own methodologies not by the
\textit{risk-sharing} framework.  It can be said that $\lambda $ is the relative
bargaining power of the Agent B, or how much the Agent A wants to enter the
contract. One can also think of $\lambda$ as the belief of how much
funding spread should be acknowledged in derivative transactions.

As we set the conventions in \cref{sec:hedging.portfolio}, $p^*$ is the amount
paid to the Agent A on top of $e _0$. This additional payment is necessary
because of default risk and funding spread. If two parties price the contracts
individually, the calculated prices may be different to each party because of
different funding spread on this model. Therefore, when the contract is made
with the initial price $e _0 + p^*$, some parties should accept a cost. Thus,
$p^*$ can be seen as the cost that is agreed by the two parties to enter the
contract, so we call $p^*$ \agrc. We also call $m^*$ optimal collateral (or
margin), and $(p^*, m^*)$ \textit{risk-sharing contract}.

Before giving the detail, we first provide motivation about the discounting
factors behind the choice of our model \cref{firstdef}. In \cref{firstdef}, the
values of the portfolios were adjusted by discounting factors. The discounting
factors are necessary for a fairness since the two agents have different funding
rates. In general, the higher default risk is, the higher funding rate
is. However, a hedging portfolio grows with respect to its funding rate (recall
\cref{hph} and \cref{hpc}). Therefore, without the discounting factors, we
penalize a party under a healthier credit condition.

One may want to put the discounting factors outside of utilities as it is a
typical choice in portfolio optimization literature. In this case, when the
portfolio processes evolve forwardly, the effect of funding rates is mixed with
risk aversion parameters in the utilities. However, the future value is purely
discounted without consideration of risk aversions, so we would again end up
with punishing or rewarding a certain party depending on risk aversions. An
argument in the same context was discussed in \cite{sircar2010utility}.

For the utilities, we will investigate two cases:
\begin{align}
     U_A(x) = x, ~~~&U_B(x) = -e^{-\gamma^Bx},\label{lin.exp}\\
    U_A(x) = -e^{-\gamma^Ax}, ~~~&U_B(x) = -e^{-\gamma^Bx}, \label{exp.exp}
\end{align}
for some $\gamma^i >0$. We choose the exponential utilities mainly for simplicity.
To use a power utility, we need for $V^i$, $i\in \{A, B\}$, to be lower
bounded. To this end, boundedness condition should be imposed to $(p, m)$, but
this makes the exposition more complicated. Moreover, an explicit form of
optimal collateral is not generally obtained under power utilities.

To solve the
\textit{risk-sharing} problem, we need different restrictions to funding spread
depending on the choice of utilities for characterizing the optimal
collateral. The restrictions are required mainly because the value functions
w.r.t \textit{variation margin} is not concave. More precisely, we will need
that $s^{B, m} = 0$ in \cref{lin.exp}, and $s^{A, m}=s^{B, m} = 0$ in
\cref{exp.exp}. The conditions on funding spread can be assumed not only when the capital
structure of a party has small leverage but also when the party achieves secured
funding for \textit{variation margin}. This situation is not common, but some
examples for the secured funding were discussed by
\cite{albanese2013restructuring}.

There is another interpretation to keep the funding condition without loss of
much generality, which is partly justified by a complete market argument. It was
shown in complete market models that an agent can guarantee that they do not
switch their position of funding state between lending and borrowing position,
depending on the structure of the payoff.  This binary nature of funding state
is related to whether payoff functions are non-increasing or non-decreasing with
respect to underlying assets. For the details, see Proposition 5.8 in
\cite{el1997backward} and refer to \cite{lee2017binary}. To streamline this
paper, we deal with cases of \cref{exp.exp}, in the main sections. For the cases
of risk-neutral agent, we report the analysis in
\cref{app:sec:risk.neutral}. Therefore, the following assumption stands
throughout the following sections except \cref{app:sec:risk.neutral}.
\begin{assumption}\label{assm.margin.rate}
\begin{enumerate}[label =(\roman*) ]
\item $s^{A, m}=s^{B, m} = 0$,
\item $U_A(x) = -e^{-\gamma^Ax}$ and $U_B(x) = -e^{-\gamma^Bx}$.
\end{enumerate}
\end{assumption}
In the next section, we represent \cref{firstdef} in a reduced form with a more
precise definition of the admissible set.  \cref{firstdef} is one type of
principal-agent problems. This problem is often called the \textit{first best}
case in typical principal-agent context. In general, it is challenging to solve
principal-agent problems because the solvability of involved equations, e.g.,
coupled FBSDEs, is not easy to obtain.  Since we also encounter a similar
difficulty as well as non-concavity, we need to modify the dynamic version of
our problem and impose some restrictions depending on the utilities.  We will
explain this point with more detail in \cref{sec:optimal.collateral}.

\subsection{Reduction of Filtration}
\label{sec:reduction} 
We start this section with introducing a long list of notations. 
The following notations are often used in this paper. For $i\in \{A, B\}$, $ t\in
[0, T]$, 
\begin{align}
  \Vb^A_t \coloneqq (B^A_t)^{-1}(V^A_t -e _{t\wedge \btau}),
  \quad~~&\Vb^B_t \coloneqq (B^B_t)^{-1}(V^B_t +e _{t\wedge \btau}), \label{xvahc}\\
    v_t \coloneqq (B^A_t)^{-1}e _t, \quad~~&c_t\coloneqq (B^A_t)^{-1}m_t,\label{vcd1}\\
    \delta _t \coloneqq v_t - c_t, \quad~~&K_t \coloneqq B^A_t(B_t^B)^{-1},\label{vcd2}\\
    \pi^i_t \coloneqq (B^i_t)^{-1} \eta^{i, S}_t\odot S_t\sigma_t,
    \quad~~&\Delta^i_t \coloneqq B_t(B^i_t)^{-1}Z_t, \label{delta}\\
  \bar{\phi}^A_t \coloneqq \pi_t^A - \Delta^A_t,
  \quad~~&  \bar{\phi}^B_t \coloneqq \pi_t^B + \Delta^B_t, \label{vf} \\
    \sigma_t\Lambda^i_t \coloneqq   (\mu_t -  R^i_t\1),\quad~~&
    b\IT \coloneqq \Lambda\IT - \Lambda_t,\label{bs}\\
    \Theta_t(\delta) \coloneqq \1_{\tau^A = t}L^A\delta^+
    &-\1_{ \tau^B = t}L^B\delta^-.
\end{align}
We give some remarks on the above notations.
\begin{remark} 
  By \cref{xvahc}, $\Vb^i$, $i \in \{A, B\}$ are (discounted) adjustment
  processes. By \cref{vcd1}, $v $ is the discounted market exposure, and $c $ is
  the discounted collateral, and by \cref{vcd2}, $\delta $ is the difference between
  the two processes. Note that $\delta=v-c=0$ means full collateralization.  By
  \cref{delta}, $\Delta^i$, $i \in \{A, B\}$, are the delta-risk of the market exposure
  adjusted by the funding rate of each party. By \cref{vf}, $\bar{\phi}^i$,
  $i \in \{A, B\}$, are the difference between the amount invested in the risky
  assets and delta risk of the \textit{clean price},i.e., $\bar{\phi}^i$ can be
  seen as the hedging error. If the Agent B does not hedge the market risk, we
  have $\bar{\phi}^B = \Delta^B$. Notice that if $b^i = 0$, then $R^i = r$, by
  \cref{bs}.
\end{remark}
We will find the projections of $\Vb^i$ onto $\dsF$, then we will deal with the
\textit{risk-sharing} problem mainly with the reduced processes.  For any
$i \in \{A, B\}$, we let $\phi^i$ denote the $\dsF$-predictable reduction of
$\bar{\phi}^i$ until $\btau$. Namely, $\phi^i$, $i \in \{A, B\}$, are
$\dsF$-predictable and
\begin{align}
 \1_{t \leq {\btau}}\bar{\phi}^i_t = \1_{t \leq {\btau}}\phi^i_t. \nonumber 
\end{align}
By It\^o's formula and \cref{vcd1}, $v $  satisfies, for $ t \in [0, T]$, 
\begin{align}
   \df v _t =& \big(-s^A_tv _t+ \Delta^A_t\Lambda_t  \big)\df t
          +\Delta^A _t\df W_t - (B^A_t)^{-1}\df \frD _t. \nonumber 
\end{align}
Note that $v $ is exogenously given. Thus, if $R^i$, $i \in \{A, B\}$, are
independent with $V^i$ and $\Vb^i$ are well-defined, then $V^i$ are also well
defined by \cref{xvahc}.
\begin{theorem}\label{theorem.reduced}
Assume $s^i$, $s^{i, m}$, $i \in \{A, B\}$, are bounded and
\begin{align}
 \SUM_{i \in \{A, B\}}\int_{0}^{T}\Big(|\delta _t|^2+|\Lambda_t^i|^2 + |\phi^i_t| + |b^i_t|^2 +
  |\Delta^i_t|^2\Big)\df  t < \infty,~~~\text{a.s.} \nonumber 
\end{align}
Then, the following processes $v^A$ and $v^B$, are well-defined:
\begin{align}
  \df v^A_t =& \big( \phi^A_t\Lambda^A_t + \Delta^A_tb^A_t+s^{A}_tv _t\big)\df t + \phi^A_t\df W_t,
  \label{smallvh}\\
  \df v^B_t =& \big(  \phi^B_t\Lambda^B_t -\Delta^B_tb^B_t-s^{B}_tK_tv _t \big)\df t + \phi^B_t\df
               W_t. \label{smallvc}
\end{align}
Moreover, assume that $R^i$, $i \in \{A, B\}$, are independent with $V^i$, and
\begin{align}
  v^A_0 =& \nu^A + p, \nonumber \\
  v^B_0 =& \nu^B - p.  \nonumber
\end{align}
Then $v^i$, $i \in \{A, B\}$, are $\dsF$-optional reductions of $\Vb^i$ until
$\btau$, i.e., $v^i$, $i \in \{H,C\}$, are $\dsF$-optional and
 $\1_{ t <\btau}\Vb^i_t =   \1_{t < \btau}v^i_t$, for any $t \geq0$.  
\begin{proof}
  It is easy to check the first assertion. To check the second part, we apply It\^o's formula to
  $(B^A_t)^{-1}V^A_t$ and this yields
\begin{align}
  \df\big((B^A_t)^{-1}V^A_t\big) =& -R^A_t(B^A_t)^{-1}V^A_t \df t +(B^A_t)^{-1}\df V^A_t
                                    \nonumber\\
  =&\1_{t \leq \btau}\pi^A_t\Lambda^A_t\df t
     + \1_{t \leq \btau}\pi^A_t \df W_t -(B^A_t)^{-1}\df
     \frC_t. \nonumber 
\end{align}
In addition, by (\ref{cashflow}) together with  $v _{\tau} =v _{\tau-}$, a.s,
\begin{align}
 (B^A_t)^{-1}\df \frC_t = \1_{t \leq \tau}(B^A_t)^{-1}\df \frD _t + \df (\1_{ \tau\leq t})
  v_{\tau-}  -\df(\1_{\tau \leq t})\Theta_{\tau}(\delta _{\tau-}).\label{lem:red.eq1}
\end{align}
Then, by combining \cref{lem:red.eq1} and \cref{lem:red.eq2}, we have
\begin{align}
  \df \Vb^A_t = &\df\big((B^A_t)^{-1}(V^A_t) -v _{t\wedge \btau}\big) \nonumber\\
  =&\1_{t \leq \btau}\big(s^{A}_tv _t+ \bar{\phi}^A_t\Lambda^A_t+\Delta^A_tb^A_t\big)\df t
     \nonumber\\
  &+\1_{t \leq \btau}\bar{\phi}_t^A\df W_t - \1_{t > \tau}(B^A_t)^{-1}\df \frD _t  -\df (\1_{ \tau \leq t})
  \big(v_{\tau-}  -\Theta_{\tau}(\delta _{\tau-}) \big).\label{lem:red.eq2} 
\end{align}
It follows that 
\begin{align}
  \df \big(\1_{t < \btau}\Vb^A_t\big)   =
  & \Vb^A_{t-}\df (\1_{t < \btau}) +
    \1_{t \leq \btau}\df \Vb^A_t -\bm{\delta}_{\btau}(\df t)\Delta \Vb^A_{\btau}\nonumber\\
  =&\Vb_{\btau-}^A\df(\1_{t < \btau}) + \1_{ t \leq \btau}\big(
     s^{A}_tv _t 
     + \bar{\phi}^A_t\Lambda^A_t + \Delta^A_tb^A_t\big)\df t
     + \1_{ t \leq \btau}\bar{\phi}^A_t\df W_t \nonumber\\
   &~ -\df(\1_{t \geq \btau})\big(v_{\tau-}  -\Theta_{\tau}(\delta _{\tau-}) \big)
     -\bm{\delta}_{\btau}(\df t)\Delta \Vb^A_{\btau} \nonumber\\
  =&\1_{ t \leq \btau}\df v^A_t -\df(\1_{t \geq \btau})\Vb^A_{\btau}
     -\df(\1_{t \geq \btau})\big(v_{\tau-}  -\Theta_{\tau}(\delta _{\tau-}) \big).\nonumber 
\end{align}
Let $\calY_t \coloneqq\1_{t < \btau}v^A_t
+ \1_{t \geq \btau}(v_{\tau-}^A-v _{\tau-}+\Theta_{\tau}(\delta _{\tau-}) )$. Again, by It\^o's formula
together with $v _{\tau} =v _{\tau-}$, a.s,
\begin{align}
  \df \calY_t =& \1_{t \leq \btau}\df v^A_t - \df(\1_{\btau \leq t}) v^A_{t-}
                 +\df (\1_{ \btau \leq t})\big(v_{\tau-}^A-v _{\tau-}+\Theta_{\tau}(\delta _{\tau-}\big) \nonumber\\
  =&\1_{t \leq \btau}\df v^A_t-\df (\1_{ \btau \leq
     t})\big(v _{\tau-}-\Theta_{\tau}(\delta _{\tau-})\big).\nonumber 
\end{align}
Thus, if $\calY_0 = \Vb^A_0$, we obtain $\calY_t = \Vb^A_t$, for any
$t\in [0, T]$. Moreover, $v^A$ is the $\dsF$-optional reduction of $\Vb^A$ and
more precisely,
\begin{align}
  \Vb^A_t = \1_{t < \btau}v^A_t + \1_{t \geq \btau}(v_{\tau-}^A-v _{\tau-}+\Theta_{\tau}(\delta _{\tau-}) ).\label{vbhd}
\end{align}
Similarly, we can  attain that
\begin{align}  
 \Vb^B_t = \1_{t < \btau}v^B_t + \1_{t \geq \btau}(v_{\tau-}^B+K_{\tau-}v _{\tau-}-K_{\tau-}\Theta_{\tau}(\delta _{\tau-}) ). \label{vbcd}
\end{align}
\end{proof}
\end{theorem}
Notice that control of $m $ is equivalent to that of $\delta $ since $e $ is given
exogenously. Thus, we solve \cref{firstdef} with respect to the two state processes
depending on $\delta $:
\begin{align}
 V^{i, p, m } = V^{i, p, \delta }.   \nonumber 
\end{align}
Moreover, we denote that $c^*\coloneqq (B^A)^{-1}m^*$ and $\delta^*\coloneqq v - c^*$.

Now, we are ready reduce the \textit{risk-sharing} problem.
Recall from \cref{firstdef} that our goal is to maximize the sum of
utilities of discounted portfolios over all $(p, \delta)\in \calA$:
\begin{align} 
  \dsE\Big[U_A\big((B^A_{\btau})^{-1}V^{A,p,\delta }_{\btau}\big)
  + \lambda U_B\big((B^B_{\btau})^{-1}V^{B,p,\delta }_{\btau}\big) \Big]. \label{goal}
\end{align}
To this end, we will represent the two terms in \cref{goal} as reduced
forms. Indeed, by \cref{lem:red}, where the integrability conditions hold
\begin{align}
  \dsE\Big[U_A\big(&(B^A_{\btau})^{-1}V^{A, p, \delta }_{\btau}\big)\Big] \nonumber\\
                   =& \dsE \Big[U_A\big(\1_{T <   \tau}v^{A, p}_T
                     +\1_{\tau \leq   T}(v^{A, p}_{\tau-}+ \Theta^{\delta }_{\tau})\big)\Big]   \nonumber\\
                   =&\dsE\bigg[G_TU_A(v^{A, p}_T) +\int_0^TG_t\Big[h^A_tU_A\big
                     (v^{A, p}_t +L^A\delta^+ _t\big) +h^B_tU_A\big(v^{A, p }_t -L^B\delta^- _t\big)\Big]\df t\bigg], \nonumber
\end{align}
and
\begin{align}
  \dsE\Big[U_B\big(&(B^B_{\btau})^{-1}V^{B, p, \delta }_{\btau}\big)\Big]\nonumber\\
  =& \dsE \Big[U_B\big(\1_{T <   \tau}v^{B, p }_T +\1_{\tau \leq   T}(v^{B, p }_{\tau-}-K_{\tau}\Theta^{\delta }_{\tau})\big)\Big]
     \nonumber\\
  =&\dsE\bigg[G_TU_B(v^{B, p }_T) +\int_0^TG_t\Big[h^A_tU_B\big
     (v^{B, p }_t -L^AK_t\delta^+\big)  +h^B_tU_B\big(v^{B, p }_t +
                     L^BK_t\delta^- _t\big)\Big]\df t\Big].
                     \nonumber
\end{align}
We define $g_t \coloneqq \1_{\delta \geq 0}g^+_t + \1_{\delta < 0}g^-_t$, where
\begin{align}
  g^+_t(v^A, v^B,  \delta) \coloneqq
  & G_t\Big[h^A_t\big(U_A(v^A + L^A\delta)
    +\lambda U_B(v^B-L^AK_t\delta)\big)+h_t^B\big(U_A(v^A) + \lambda U_B(v^B )\big)\Big] \nonumber\\\
  g^-_t( v^A, v^B, \delta) \coloneqq
  & G_t\Big[h_t^B\big(U_A(v^A +L^B\delta) + \lambda U_B(v^B - L^BK_t\delta)\big) 
    +h_t^A\big(U_A(v^A) + \lambda U_B(v^B) \big)\Big].\nonumber 
\end{align}
For the above reduction to be valid, we assume the following integrbility condition:
\begin{align}
  \SUM_{i \in \{A, B\}}\bigg[|U_i(v^i_T)| + \int_{0}^{T}|U_i(v^i_t)| \df t  \bigg] <\infty,
  \label{condition.integrability} 
\end{align}
and we define the admissible set of collateral $\calD$ for a given Borel set $A
\subseteq \dsR$ as follows:
\begin{definition}
  $\delta  \in \calD$, if $\delta  \in \dsH^2_T$ and
\begin{enumerate}[label=(\roman*)]
\item $\delta  \in A$, $\dpdt$,
\item $\dsE \big[\int_{0}^{T}\big\vert g_t(v^{A, p}_t, v^{B, p }_t, \delta _t)\big\vert\df t\big]<\infty.$
\end{enumerate}
\end{definition}
Then, the \textit{risk-sharing} problem can be rewritten as
\begin{align}
  \max_{(p, \delta )\in \calA}\dsE
  \bigg[G_TU_A(v^{A, p}_T) + \lambda G_TU_B(v^{B, p}_T) +
    \int_{0}^{T}g_t(v^{A, p}_t, v^{B, p}_t, \delta _t)\df t\bigg],
  \label{pre}
\end{align}
where $\calA = \dsR \times \calD$ and  $v^i$, $i \in \{A, B\}$, are defined in
(\ref{xvahc})-(\ref{bs}), (\ref{smallvh}), and (\ref{smallvc}). Note that $v^i$
does not depend on $\delta$ since we assume $s^{A, m} = s^{B, m}=0$. When the Agent A
is risk-neutral, we only need that $s^{B, m}=0$, and in this case $\calD$ should be
defined in a slightly different way. We will discuss this with more details in
\cref{app:sec:risk.neutral}. 
\begin{remark}\label{rem.difficulty} 
\begin{enumerate}[label=(\roman*)]
\item The \textit{risk-sharing} framework can be thought of as a two-agent problem. 
  Since there is no party who can solely decide the contract, the
  mathematical structure of our \textit{risk-sharing} problem is different from
  that of typical principal-agent problems. For example, for the Agent A, in both
  perspectives of funding impacts and loss given defaults, posting collateral to
  the Agent B is not beneficial. Say, we consider $A$ as an agent,
  subject to $B$ as a principal. Then, if we  solve the agent problem first,
  e.g., as in \cite{cvitanic2018dynamic}, it always gives the trivial solution
  $\delta^* = v  -c^*=\infty$.
\item  One may want to take stochastic calculus of variation as in
\cite{cvit2013contract}. However, note that $g$ is piece-wise concave in
$\delta$. Even when $g$ is concave, it may not be differentiable. Therefore, if we
take stochastic calculus of variation, we will face a very challenging FBSDE
with a discontinuous coefficient in the drift of (forward) SDE. A similar case
was dealt with by \cite{chen2018forward}. However, in our case, we encounter
multi-dimensional FBSDE with a degenerate volatility and unbounded
coefficients. The solvability of such FBSDE is beyond the scope of this paper
and we leave it as future research. Instead, in this paper, we impose some
conditions on funding cost/benefit in delivering the collateral, $s^{i, m}$, $i
\in \{A, B\}$, depending on the utilities, and use verification argument. 
\end{enumerate}
\end{remark}

\section{Optimal Collateral}
\label{sec:optimal.collateral}
In this section, we characterize the optimal collateral in the
\textit{risk-sharing} problem by using martingale optimality principle. Then we
argue by verification that the characterized collateral is indeed an optimal
solution.  First, we solve the problem with respect to only \textit{variation
  margin} $\delta $, with a fixed initial price $p \in\dsR$:
\begin{align}
 \max_{\delta \in \calD}\dsE\bigg[G_TU_A(v^{A, p, \delta }_T) + \lambda G_TU_B(v^{B, p, \delta }_T) +
  \int_{0}^{T}g_t(v^{A, p, \delta }_t, v^{B, p, \delta }_t,  \delta _t)\df t\bigg]. \label{sold}
\end{align}
Then, the \agrc\ $p^*$ will be found with the given optimal \textit{variation
  margin} $\delta^*$.  However, mainly because of the non-concave property of our
problem addressed in \cref{rem.difficulty}, we need to impose some restrictions
to the funding spread in delivering collateral for the characterization. Recall
that we consider two cases: a risk-neutral Agent A and risk-averse the Agent B
investing their capital with a small leverage so that $s^{B,m} =0$, and two risk
averse parties with $s^{B,m}=s^{A, m}=0$. As we mentioned, the mathematical
analysis for a risk-neutral agent is deferred to \cref{app:sec:risk.neutral}.
In what follows, we first derive an optimal collateral for both cases,
then we give financial interpretations later. The two most notable features are
the weak dependence with default intensities and relationship with the full
margin requirement. The discussion about the relationship between the optimal
collateral and margin requirement is an important part of this paper.This
funding condition is not necessary for finding $p^*$ if $\delta$ is not a control
variable, e.g. $A = \{\delta_0\}$ for some $\delta_0 \in \dsR$.
  
We define a dynamic version of (\ref{pre}) and use martingale optimality
principle (MOP) as in \cite{jiao2011optimal}. To this end, we define a set of
controls which coincide with a given $\varepsilon \in \calD$ up to a certain time
$t \leq T$. We denote the set by $\calD(t, \varepsilon)$, i.e., for
$\varepsilon \in \calD$,
$\calD(t, \varepsilon)\coloneqq\big\{\delta  \in \calD\big| \delta _{. \wedge t} = \varepsilon_{\cdot \wedge t}\big\}$.
Now, define the dynamic version of (\ref{sold}) as
\begin{align}
  J^\varepsilon_t(p) \coloneqq \esssup_{\delta \in \calD(t, \varepsilon)}\dsE\bigg[G_TU_A(v^{A, p}_T)
  + \lambda G_TU_B(v^{B, p}_T) + \int_{t}^{T}g_s(v^{A, p}_s, v^{B, p}_s,
  \varepsilon_s)\df s \bigg|\calF_t\bigg]. \label{dpre2}
\end{align}
Then we characterize the optimal collateral by using martingale optimality
principle. By MOP, $(J^\varepsilon_t)_{0\leq t\leq T}$ is chosen as a c\`adl\`ag version such
that  for any $ \varepsilon \in \calD$,
\begin{align}
  \Big\{J^\varepsilon_t + \int_{0}^{t}g_s(v^{A}_s, v^B_s, \varepsilon_s) \df s\Big\}_{0 \leq t \leq T}\nonumber 
\end{align}
is a $(\dsP, \dsF)$-supermartingale. Moreover, for the optimal collateral $\delta^*$ for
$J_0$,
\begin{align}
  \Big\{J^{\delta^*}_t + \int_{0}^{t}g_s(v^A_s, v^B_s,\delta^*_s)\df s\df s\Big\}_{0 \leq t \leq T}  \nonumber 
\end{align}
is a $(\dsP, \dsF)$-martingale. When the admissibility is guaranteed, a
solution to (\ref{sold}) can be found by verification. 

Before moving on, to represent the optimal collateral by one stochastic process,
we define a process
$(X_t)_{t\geq0}$ such that
\begin{align}
  U_A(X_t) \coloneqq \frac{U_A(v^{A }_t)}{-U_B\big(v^{B }_t -v^{B}_0\big)}
  =-\exp{\bigg[-\gamma^A\Big(v^{A }_t
  - \gratio (v^{B }_t -v^{B}_0)\Big)\bigg]}. \label{defx}
\end{align}
Namely, $X_t = v^{A }_t - (\gamma^B/\gamma^A)(v^{B}_t -v^{B}_0)$. More precisely, $X$ is given by
\begin{align}\label{def.X}
    X_t =&v_0^A  + \int_0^t\Big[s_tv _t+\phi^A_t\Lambda^A_t - \gratio \phi^B_t\Lambda^B_t + \Delta^A_tb^A_t+\gratio
                 \Delta^B_tb^B_t\Big]\df s + \int_{0}^{t}\phi_s\df W_s,
\end{align}
Then, (\ref{sold}) will be represented w.r.t $X$, and 
where $\phi_t \coloneqq \phi^A_t -(\gamma^B/\gamma^A)\phi^B_t$ and
$ s_t\coloneqq s^{A}_t + (\gamma^B/\gamma^A) s^{B}_tK_t$. 
\begin{theorem}\label{theorem.ee} Assume that the integrability condition
  \cref{condition.integrability} hold. Define
  \begin{align}
  \delta^*_t(p, x) \coloneqq& \argmax_{\delta \in A}\big\{U_A(x)\psi^A_t( \delta)
                  +\lambda U_B(v_0^{B, p}) \psi^B_t(\delta)\big\}, \label{sol.ee}\\
  \psi^A_t( \delta) \coloneqq &-h^A_tU_A(L^A\delta^+) -  h^B_tU_A(-L^B\delta^-), \label{psih}\\
  \psi^B_t( \delta) \coloneqq &-h^A_tU_B(-L^AK_t\delta^+)- h_t^BU_B(L^BK_t\delta^-). \label{psic}
\end{align}
  If $\csolt \in \calD$, then $\csolt$ is a solution of (\ref{sold}). Moreover,
\begin{align}
  J_0=\dsE\bigg[ \beta_T\big[U_A(X_T) + \lambda U_B(\nu^B-p)\big] + \int_{0}^{T}\fh_t(p, X_t) \df t \bigg], 
  \label{theorem.ee.result}
\end{align}
where $ \beta_t\coloneqq-G_tU_B\big(v_t^{B} -v^{B}_0\big)$ and
\begin{align}
  \fh_t(p, x)\coloneqq
  \beta_t\big[U_A(x)\psi^A_t( \delta^*_t(p, x)) +\lambda U_B(\nu^B-p) \psi^B_t(\delta^*_t(p, x))\big].\label{def.fh}
\end{align}
\begin{proof}
  For $\varepsilon \in \calD$, define
  $\xi^\varepsilon_t\coloneqq J_t + \int_{0}^{t}g_s(v^{A}_s, v^{B}_s, \varepsilon_s)\df s$.
  Notice that $J_t$ is independent of $\varepsilon\in \calD$ and by \cref{psih} and \cref{psic},  
\begin{align}
  -\big[G_tU_B(v^{B}_t - v^B_0)\big]^{-1} g_t(v^{A}_t,  v^{B}_t, \varepsilon_t)
  =&U_A(X_t)\psi^A_t(\varepsilon_t) +\lambda U_B(v_0^B) \psi^B_t(\varepsilon_t). \nonumber 
\end{align}
Therefore, for any $\varepsilon \in \calD$, $ \xi^\varepsilon - \xi^{\csolt} $  is a
$(\dsP, \dsF)$-supermartingale. Moreover, for any $\epsilon \in \dsD$
\begin{align}
  \dsE\big[\xi^\epsilon_T - \xi^{\delta^*(p, X)}_T\big] \leq \dsE \big[\xi^\epsilon_0 - \xi^{\delta^*(p, X)}_0\big]=0.
\end{align}
Thus, (\ref{theorem.ee.result}) is obtained
where the admissibility of $\csolt$ is guaranteed.
\end{proof}
\end{theorem}
To find the explicit form of $\csolt$, we consider $A = \dsR$ and represent
(\ref{sol.ee}) as
\begin{align}
  \delta^*_t(p,x) \coloneqq \argmax_{\delta \in A}
  \big(\1_{\delta  <0}f^-(t, p, x, \delta)
  +\1_{\delta  \geq0}f^+(t, p, x, \delta) \big),    \nonumber
\end{align}
for some functions $f^-, ~f^+$. Then, $f^i$, $i \in \{-, +\}$ are continuously
differentiable in $\delta$ and for any $(t, p, x)$, there exist $I^i_t( p, x)$ such
that
\begin{align}
  \partial_\delta f^i(t, p, x, I^i_t(p, x))=0. 
\end{align}
Then, $\csolt$ can be attained at $I^-, ~I^+$, and zero. We can easily see that 
\begin{align}
  f^-(t, p, x, \delta) \coloneqq
  & h_t^B\big[U_A(x+L^B\delta) + \lambda U_B(\nu^B - p-L^BK_t\delta)\big]\nonumber\\
  &\quad+h^A_t\big[U_A(x) + \lambda U_B(\nu^B - p)\big],\label{def:f-}\\
    f^+(t, p, x, \delta) \coloneqq
  & h_t^A\big[U_A(x +L^A\delta) + \lambda U_B(\nu^B-p-L^AK_t\delta)\big] \nonumber\\
  & \quad+h^B_t\big[U_A(x ) + \lambda U_B(\nu^B - p - L^BK_t)\big].\label{def:f+}
\end{align}
Therefore, we obtain that
\begin{align}
    I^-_t( p, x) \coloneqq
  & \frac{\gamma^B\nu^B- \gamma^B p -\gamma^Ax 
    - \ln{\big( \lambda K_t \gratio}\big)}{L^B(\gamma^BK_t+\gamma^A)}, \label{def.im.ee}\\
 I^+_t( p, x) \coloneqq
  &\frac{\gamma^B\nu^B- \gamma^B p -\gamma^Ax 
    - \ln{\big( \lambda K_t \gratio}\big)}{L^A(\gamma^BK_t+\gamma^A)}. \label{def.ip.ee}
\end{align}
The exact form of $\delta^*$ can be obtained by characterizing the region
\begin{align}
\Big\{\max_{\dsR} f^- > \max_{\dsR} f^+\Big\}. \nonumber     
\end{align}
The calculation of the region is a straightforward but tedious; see, e.g., 
\cite{carassus2009portfolio}. We only obtain the exact form for a simple case
which will be seen later.
We complete \cref{theorem.ee} by the next lemma. The proof is reported in \cref{app:sec:proof}.
\begin{lemma}\label{lemma.ee}
Let  $A = \dsR$, and assume $(e_t)_{t\geq0}, (Z_t)_{t\geq0}, (\pi^i_t)_{t\geq0}, ~i \in \{A,
B\}$, are bounded. Then $\csolt \in \calD$ and \cref{condition.integrability} hold.
\end{lemma}
 
\begin{example} Consider an Agent A who
  hedges delta-risk and an Agent B who does not hedge, i.e., 
  $\pi^A = \Delta^A, ~\pi^B = 0$. 
They enter a bond contract that is paid by the Agent A, namely,
$\frD  = \1_{\llbracket T, \infty  \rrbracket}$. 
We assume that OIS rate $(r_t)_{t\geq0}$ follows the next SDE:
\begin{align}
 \df r_t = k(\theta - r_t)\df t + \rho \sqrt{r_t}\df W^{\dsQ}_t,\nonumber 
\end{align}
for some $k, \theta, \rho \in \dsR$ and a risk-neutral measure $\dsQ$. Moreover, we assume
that $h^i$, $i\in \{A, B\}$, are bounded and $s^B = s^{B,m}= 0$. Then, by Clark-Ocone formula, for
$t \geq 0$,
\begin{align}
  Z _t =& -\rho \sqrt{r_t}A^2(t, T)B^{-1}_te _t,\nonumber 
\end{align}
where
\begin{align}
  e _t =& A^1(t, T)e^{-r_tA^2(t, T)},\nonumber \\
  A^1(t, T) \coloneqq & \Big(\frac{2ae^{(a+k)(T-t)/2}}{2a +
                (a+k)(e^{a(T-t)}-1)}\Big)^{2k\theta /\rho^2},\nonumber \\
  A^2(t, T) \coloneqq &\frac{2(e^{a(T-t)} - 1)}{2a + (a + k)(e^{a(T-t)} - 1)},\nonumber \\
  a \coloneqq& \sqrt{k^2 + 2 \rho^2}. \nonumber 
\end{align}
Since $r >0$, $Z $ is bounded. Hence, all conditions in \cref{lemma.ee} are satisfied.
\end{example}

Now, we are ready to discuss the financial interpretation of the optimal
collateral. In the next section, the financial meanings of
(\ref{def.im.ee})-(\ref{def.ip.ee}) and the
relationship with the margin requirement will be discussed.
\subsection{Analysis of Collateral}
\label{sec:analysis} 
In this section, we provide financial interpretations of the optimal collateral
derived in the previous sections. Collateral is posted for default risk. In our
model, there are two main components in default risk: intensities and loss
rates. We first discuss a weak dependence between the optimal collateral and
default intensities. We begin with giving the explicit form of $\delta^*$ in the
following lemma. We report the proof in \cref{app:sec:proof}.
\begin{lemma}\label{lem:col} Assume $A = \dsR$. Then, $\delta^*_t(p, x)$ is given by
\begin{align}
  \delta^*_t(p, x) =&
   (0 \vee I^+_t( p, x) )  + (0 \wedge I^-_t( p, x) ),~~\text{where} \label{opt:col0}\\
    I^-_t( p, x) \coloneqq& \frac{-\gamma^Ax - \gamma^B p}{L^B(\gamma^BK_t+\gamma^A)} + \frac{\gamma^B\nu^B
                  - \ln{\big( \lambda K_t\gratio}\big)}{L^B(\gamma^BK_t+\gamma^A)}, \label{iminus0}\\
  I^+_t( p, x) \coloneqq& \frac{-\gamma^Ax - \gamma^B p}{L^A(\gamma^BK_t+\gamma^A)} + \frac{\gamma^B\nu^B
                  - \ln{\big( \lambda K_t\gratio}\big)}{L^A(\gamma^BK_t+\gamma^A)}. \label{iplus0}
\end{align}
\end{lemma}
Note from \cref{opt:col0}-\cref{iplus0} that the optimal collateral depends only
on the loss rates $L^i$ not on default intensities $h^i$, which is a rather
natural consequence. Collateral is required for loss given default not for the
default itself. Put differently, collateral is about how much loss would be
inflicted at default and not about how likely default occurs. Recalling
$\delta^* = v - c^*$ and observing (\ref{iminus0}) and (\ref{iplus0}),
the magnitude of the optimal \textit{variation margin} $c^*$ increases as $L^i$,
$i \in \{A, B\}$, increase.

We discuss the effect of loss rates with more details.
By (\ref{opt:col0}), when $\csolt \leq 0$, $I^-(p, X) = \csolt$. In this case, as
$L^B$ increases, $c^* = v -\csolt$ decrease sbecause of the increased  average
loss of collateral posted to the Agent B. On the other hand, when
$\csolt \geq0$, the optimal collateral $c^* = v -\csolt$, is independent of
$L^B$ and increases w.r.t $L^A$. Again, this is because the high loss rate makes
it risky for the Agent B to post collateral to the Agent A.

The relationships with $p$ and  $\lambda$ are self-explanatory. If
the contract starts from giving a high price $p$, to the Agent A at initiation
of the contract, the Agent A needs to post more collateral in return. Moreover,
the higher $\lambda$  is, i.e., the strong bargaining power the Agent B has, the more
the Agent A should post more collateral.

In addition, recalling $X_t = v^A_t - (\gamma^B/\gamma^A)(v^B_t - v_0^B)$,
it seems that \cref{opt:col0} suggests that optimal collateral ratio should be
decided by the relative performance of each party. It is not obviously
applicable in practice. However, we can use $X$ to derive an interesting
interpretation from the full margin requirement of Basel \Rom{3}, which will be
discussed in the next section. 
\begin{remark}
From (\ref{opt:col0})-(\ref{iplus0}), the major factor for
collateral is the loss rate.  In practice, loss rates are often chosen as
$0.6$ regardless of entities. Our model together with the practice on loss
rates partly explains the margin requirement applied to all banks.
\end{remark}
\subsection{Analysis of the Full Margin Requirement}
\label{sec:margin}
In this section, we interpret the meaning behind the inter-dealer market
convention that is required by Basel \Rom{3}. It can be understood that  the inter-dealer
convention is $\delta^*(p, X) = 0 $.
By (\ref{def.im.ee}) and (\ref{def.ip.ee}), the full margin convention requires
that 
\begin{align}
  X_t + \frac{1}{\gamma^A}(s^A-s^B)t = \frac{\gamma^B}{\gamma^A}(\nu^B - p) - \frac{1}{\gamma^A}\ln{\bigg(
  \frac{\lambda\gamma^B}{\gamma^A}\bigg)}, ~~\dpdt. \label{xconstant}
\end{align}
Therefore, since $\{X_t + (\gamma^A)^{-1}\int_{0}^{t}(s^A_s-s^B_s)\df s\}_{t\geq0}$ should be
constant,  it is necessary that 
 $\phi = 0$. If two parties' hedging strategies are chosen independently
of each other, by (\ref{def.X}), $\phi = 0$ may mean
 $\phi^A = \phi^B =0$.
It follows that $\pi^A = \Delta^A$ and $\pi^B = -\Delta^B$. In other words, the two
parties should  hedge the delta-risk of market exposure. In addition, together
with \cref{def.X}, this constant condition implies that 
\begin{align}
  \bigg(s^A + \frac{\gamma^B}{\gamma^A}s^BK_t\bigg) v
  +\Delta^A b^A +\gratio \Delta^B b^B +\frac{s^A-s^B}{\gamma^A}=0, ~\dpdt. \label{implication.cond}
\end{align}
For \cref{implication.cond} to hold with arbitrary $\Delta^A$ and $\Delta^B$, we should
have $s^A = s^B=0$. Therefore, for the market convention to be optimal, the
following two conditions are necessary:
\begin{itemize}
\item both agents hedge the delta-risk of \textit{clean price},
\item funding spreads are not transferred to each party.
\end{itemize}
The second item seems like an expected result because the condition
$\delta^* = 0$ inherently considers two parties whose earnings from the margin is
symmetric. If one can make a profit or suffer a loss by margin process,
$\delta^* = 0$ may not be optimal. A debate is still underway whether funding spread
should be recouped from counterparties and how to handle the accounting; see,
e.g., \cite{hull2012fva, hull2012fva2, castagna2012yes, andersen2019funding,
  albanese2014accounting}. Indeed, in frictionless markets, choices of funding
are separated with pricing  as MM theorem
properly applies. However, with
frictional distress costs, shareholders' decision can depend on the choices of
funding. In such cases, the margin requirement is not optimal anymore.
Therefore, the second condition on funding transfer can be understood that the
margin requirement of Basel \Rom{3} inherently considers frictionless financial
markets.

In the next section, we will derive a maximum principle of $p^*$ for \cref{pre}.
Mainly because of an issue from non-concavity, for finding the optimal pair
$(p^*, \delta^*)$, we need either $s^{B, m} = s^{A, m} = 0$ or
$A = \{\delta_0\}$, for some $\delta_0\in \dsR$. The second condition means that the
\textit{variation margin} $c$ is fixed as a given process.
 
\section{Optimal Initial Prices}
\label{sec:optimal.price}
Throughout this section, the conditions in \cref{lemma.ee} are assumed so that the admissibility 
is obtained. The next maximum principle for $p^*$ is basically a first order
condition. First, we consider the case that $\delta$ is not a control variable,
i.e. $A = \{\delta_0\}$, for some $\delta_0\in \dsR$. In previous sections, we have assumed
that $s^{A, m} = s^{B, m}=0$. However, when $A$ is singleton, we do not need
the condition on margin rate.  
\begin{theorem} \label{thm:ps} Assume $A = \{\delta_0\}$, for some $\delta_0\in \dsR$. Therefore,
  $ \delta^* = \delta_0$. Let
  $X^* \coloneqq X^{p^*, \delta^*}$,
  $f^*_\cdot \coloneqq\fh_\cdot(p^*, X^*_\cdot)$, and for given $t\leq T$, define
  $Q_t\in \dsR^2$ as the set that $f^*_t(\cdot)$ is not differentiable. Assume
\begin{align}
  \1_{(p^*, X^*) \in Q} = 0, ~~\dpdt, \label{nullQ}
\end{align}
i.e., $(p*, X^*)$ does not fall in the non-differentiable set of $\fh$, $\dpdt$. Moreover, assume
\begin{align}
\dsE\Bigg[\beta_TU'_A(X^*_T)- &\beta_T\lambda U'_B(\nu^B-p^*)\nonumber\\
  &+\int_{0}^{T}\1_{(p^*, X^*_t)\notin Q_t}
  \big(\partial_p \fh_t(p^*, X^*_t)+\partial_x \fh_t(p^*, X^*_t)\big)\df t \Bigg]=0.\label{mpp}
\end{align}
Then $p^*$ is the optimal initial price. 
\begin{proof} Notice that $f^*_t( \cdot)$ is concave for any $t \in [0, T]$, so is differentiable
  a.e. The maximum principle (\ref{mpp}), is basically a first order
  condition. We only need to check whether $(p^*, X^*)$ is not absorbed in
  $Q$. Since we assume  $A = \{\delta_0\}$, for some $\delta_0\in \dsR$, $\delta^*$ does not 
  depend on $(p, X)$. We let, for any process $\varphi$,
  $\varphi^* \coloneqq \varphi^{p^*}$ and, for arbitrary $p \in \dsR$,
  $\Delta \varphi^* \coloneqq \varphi^p - \varphi^*$. Then, 
\begin{align}
  \dsE\big[\Delta Y^*_0\big]
  =& \dsE\big[\beta_T \big(U_A(X^p_T) - U_A(X^*_T)\big)
 + \beta_T\lambda \big(U_B(\nu^B-p) - U_B(\nu^B-p^*)\big) +\int_{0}^{T}\Delta f^*_t\df t\big]\nonumber
  \\
  \leq&\dsE\big[\beta_TU'_A(X^*_T)\Delta X^*_T- \beta_T\lambda U'_B(\nu^B-p^*)\Delta p^*
     +\int_{0}^{T}\Delta f^*_t \df t \big]\nonumber \\
  \leq &\dsE\big[\beta_TU'_A(X^*_T)\Delta X^*_T- \beta_T\lambda U'_B(\nu^B-p^*)\Delta p^*\big]\nonumber \\
   &+\dsE\bigg[\int_{0}^{T}\1_{(p^*, X^*_t)\notin Q_t}\big(\partial_x \fh_t(p^*, X^*_t)\Delta X^*_t
     +\partial_p \fh(p^*, X^*_t)\Delta p^*\big)\df t \bigg].\nonumber 
\end{align}
The last inequality is obtained by concavity of $f^*$ and (\ref{nullQ}). Notice
that $\Delta X^*_t = \Delta p^*$, for any $t\in [0, T]$. Therefore, by \cref{mpp}, we have
\begin{align}
\dsE\big[Y^p_0 - Y^*_0\big]
  \leq&\Delta p^*\dsE\big[\beta_TU'_A(X^*_T)- \beta_T\lambda U'_B(\nu^B-p^*)\big]\nonumber \\
   &+\Delta p^*\dsE\bigg[\int_{0}^{T}\1_{(p^*, X^*_t)\notin Q_t}\big(\partial_p \fh_t(p^*, X^*_t)
     +\partial_x \fh(p^*, X^*_t)\big)\df t \bigg]\nonumber \\
  =&0.\nonumber 
\end{align}
\end{proof}
\end{theorem}
When we control $(p, \delta)$ together, two conditions on the funding spread,
$s^{A, m}=s^{B, m}=0$, are required. The proof is analogous to that of \cref{thm:ps}.  
\begin{proposition} Assume $s^{A, m}=s^{B, m}=0$. Let $X^* \coloneqq X^{p^*}$,
  $f^*_\cdot \coloneqq\fh_\cdot(p^*, X^*_\cdot)$, and for given $t \leq T$, define
  $Q_t\in \dsR^2$ as the set that $f^*_t(\cdot)$ is not differentiable. Assume
\begin{align}
  \1_{(p^*, X^*) \in Q} = 0, ~~\dpdt, \nonumber
\end{align}
i.e., $(p*, X^*)$ does not fall in the non-differentiable set of $\fh$, $\dpdt$. Moreover, assume
\begin{align}
\dsE\Bigg[\beta_TU'_A(X^*_T)- \beta_T\lambda U'_B(\nu^B-p^*)
  +\int_{0}^{T}\1_{(p^*, X^*_t)\notin Q_t}
  \big(\partial_p \fh_t(p^*, X^*_t)+\partial_x \fh_t(p^*, X^*_t)\big)\df t \Bigg]=0. \nonumber 
\end{align}
Then $(p^*, \delta^*)$ is the \textit{risk-sharing contract}. 
\end{proposition}
We deal with examples in the next section.

\section{Examples}
\label{sec:example}
In \cref{sec:margin}, it was shown that delta-hedge of \textit{clean price} and
the absence of market frictions are necessary for the full margin requirement to
be optimal. We first derive the \textit{risk-sharing} contract given the conditions.
\begin{example}
Assume $s^{i, m}=s^i=\phi^i = 0$, $i\in \{A, B\}$, $A = \dsR$. Therefore,
$X^p = \nu^A + p$, and $\beta = G$. We will check that $( p^*, \delta^*) = (\ph, 0)$ where
\begin{align}
 \ph \coloneqq \frac{\gamma^B\nu^B - \gamma^A\nu^A}{\gamma^B+\gamma^A} - \frac{1}{\gamma^B+\gamma^A}\ln{\bigg(
  \frac{\lambda\gamma^B}{\gamma^A}\bigg)}. \label{def.ph}
\end{align}
By \cref{iminus0}-\cref{iplus0}, we have
$I^+_t(X^p_t, p) = (L^A)^{-1}(\ph - p), ~I^-_t(X^p_t, p) = (L^B)^{-1}(\ph -
p)$, where $\ph$ is defined as \cref{def.ph}. Thus, by taking $p = \ph$, we
recover the full margin convention: $\delta^* = 0$. In addition, 
\begin{align}
  X^{\ph} + L^AI^+ =&\nu^A+\ph,\nonumber \\
  \nu^B-p+L^BI^- =& \nu^B-\ph.\nonumber 
\end{align}
Therefore, by \cref{def.fh}, $\fh_t$ is differentiable at $(\ph, X^{\ph})$, i.e.,  for $ t\in
[0, T]$, $Q_t = \emptyset$. Moreover, by straight forward calculation, $\partial_p \fh_t(p, x)
= -\partial_x \ft_t(p, x) $, and 
\begin{align}
 U'_A(\nu^A+\ph) =\lambda U'_B(\nu^B-\ph). \nonumber 
\end{align}
Therefore, we obtain $(p^*, \delta^*) = (\ph, 0)$. Note that when
$\gamma^B = \gamma^A$, $\nu^A = \nu^B = 0$, $\lambda=1$, then $\ph=0$. Thus, in this case,
$(p^*, \delta^*) = (0,0)$.
\end{example}
If we take $\gamma^B = \gamma^A=1$ and $\nu^B = \nu^A = 0$, \cref{def.ph} is reduced to 
$\ph = -\ln{(\lambda)}/2$.  In particular, when the two parties have the same
negotiation power, i.e., $\lambda=1$, we have $\ph =0$.  It can be said that
$\ph$ represents the amount of adjustment by agents' preference and negotiation
power, which are non-observable information in markets. Since it is hard for
both parties to agree on such parameters. In addition, it is 
notable that $\ph$ does not depend on $h^i$, $L^i$, $i \in \{A, B\}$, because the
price is mathematically derived from fully collateralized contracts. If one
party does not hedge the delta-risk, we cannot have an explicit solution for
$\ph$, so we discuss only the existence of $p^*$ satisfying \cref{mpp}.
\begin{example}
Assume $s^{i, m}=0$, and  $\pi^A = \Delta^A$, $\pi^B = 0$, i.e.,
$\phi^A=0$, $\phi^B = \Delta^B$. We consider constant default intensities and, without
loss of generality, assume that $\gamma^A = \gamma^B$, $L^A = L^B = 0.5$,
$\lambda =1$, and $\nu^A =\nu^B= 0$. Therefore, for $ t \in [0, T]$,
\begin{align}
  &X^p_t = p - \int_{0}^{t}\Big[\Big(s^A_s + \frac{\gamma^B}{\gamma^A}s^B_s\Big)v _s
    + \Delta^B_s(b^B_s-\Lambda^B_s) + \Delta^A_sb^A_s\Big]\df s 
    -\int_0^t\Delta^B_s \df W_s,\nonumber \\
&I^i_t(p, X^p_t) = -X^p_t - p, ~i \in \{-, +\}. \nonumber 
\end{align}
Since $\gamma^A = \gamma^B$, we denote
$  U\coloneqq U_A = U_B  $.
By straightforward calculation, we can
check that $Q_t = \emptyset$, for $t \leq T$, and  
\begin{align}
  \partial_p\fh_t(p, X^p_t)+\partial_x\fh_t(p, X^p_t)
  &= \left\lbrace
  \begin{array}{r@{}l}
    &-\gamma h^B\beta_t \big( U(X^p_t) - U(-p)\big), ~~-X^p_t -p \geq0, \\
    &-\gamma h^A\beta_t \big( U(X^p_t) - U(-p)\big), ~~-X^p_t -p <0. 
 \end{array}
  \right.\nonumber 
\end{align}
Recall that $X^p_T$ increases as $p$ increases. Therefore, both
$\partial_p \fh + \partial_x \fh$ and $[U'(X_T) - U'( -p)]$ decrease w.r.t $p$. Moreover, both
terms tend to $\infty$ (resp. $-\infty$) as $p\rightarrow -\infty$ (resp.
$p\rightarrow \infty)$. Thus, there exists $p \in \dsR$ satisfying \cref{mpp}. Once
$p^*$ is obtained, $\delta^*$ can be found as well, but in this case,
$(p^*, \delta^*)$ may not be $(\ph, 0)$, i.e., full collateralization may not be
optimal.
\end{example}
\section{Conclusion}
In this paper, we introduced a new \textit{risk-sharing} framework to understand
how two parties enter bilateral contracts with the presence of entity-specific
information such as default risk and funding spread. Based on our model, we can
explain why banks buy Treasury bonds that return less than their funding
rate. The analysis of the optimal collateral in the \textit{risk-sharing}
framework interprets the meaning behind the margin requirement in Basel \Rom{3}:
two parties hedge delta risk of \textit{clean price} and funding spread is not
considered in derivative prices. Note that the full collateralization is really
optimal in frictionless financial markets, which is an inherent assumption in
Basel \Rom{3}.  It is possible that this conclusion can change if we include
\textit{gap risk}, KVA, and hedging strategies are also control variables. We
leave such analysis as a further research topic.  
\appendix
\begin{appendices}
\section{An Auxiliary Lemma}
\label{sec:lemma}  
The next lemma is borrowed from \cite{bielecki2008pricing} and often used in
this paper. 
\begin{lemma}\label{lem:red} Let $i \in \{A, B\}$.
\begin{enumerate}[label=(\roman*)]
\item Let $U$ be an $\calF_s$-measurable, integrable random variable for some $s
  \geq0$. Then, for any $t \leq s$,
  \begin{align}
    \dsE(\1_{s <\tau}U\vert \calG_t) =& \1_{t < \tau}G_t^{-1}
    \dsE (G_sU \vert\calF_t). \nonumber
  \end{align}
\item Let $(U_t)_{t \geq0}$ be a real-valued, $\dsF$-predictable process and
  $\dsE\vert U_{\btau} \vert < \infty$. Then,
  \begin{align}
    \dsE(\1_{\tau = \tau^i \leq T}U_{\tau}\vert \calG_t) =
    \1_{t< \tau}G_t^{-1}\dsE\Big( \int_{t}^{T}h^i_s G_sU_s\df s \Big\vert \calF_t\Big). \nonumber
  \end{align}
\end{enumerate}
\end{lemma}
\section{Spaces of Random Variables and Stochastic Processes}
\label{sec:spaces}
In this paper, we denote spaces of random variables and stochastic processes as follows.
\begin{definition} Let $m \in \mathbb{N}$ and $p \geq 2$.
\begin{itemize}
\item $\dsL^p_T$: the set of  all $\calF_T$-measurable random variables $\xi$, such that
 \begin{align}
   \Vert\xi \Vert_p\coloneqq \dsE[|\xi|^p]^{\frac{1}{p}}<\infty. \nonumber 
  \end{align}
\item $\dsS^p_T$: the set of  all real valued,
  $\dsF$-adapted, c\`adl\`ag\footnote{Right continuous
    and left limit.} processes $(U_t)_{t\geq0}$, such that
  \begin{align}
    \Vert U \Vert_{\dsS_T^p} \coloneqq \dsE\big(\sup\limits_{t \leq T}\vert U_t\vert^p\big)^{\frac{1}{p}}< \infty. \nonumber
  \end{align}
\item $\dsH^{p, m}_T$: the set of all $\dsR^m$-valued, $\dsF$-predictable
   processes $(U_t)_{t\geq0}$, such that
  \begin{align}
    \Vert U \Vert_{\dsH^p_T} \coloneqq \dsE\Big(\int_0^T\big\vert U_t \big\vert^p\df t\Big)^{\frac{1}{p}} < \infty. \nonumber
  \end{align}
\item $\dsH^{p, m}_{T, loc}$ : the set of all $\dsR^m$-valued, $\dsF$-predictable
   processes $(U_t)_{t\geq0}$, such that
  \begin{align}
    \int_0^T\big\vert U_t \big\vert^p\df t< \infty, \quad \text{a.s.} \nonumber
  \end{align}
\end{itemize}
\end{definition}
When $d=1$, we denote  $\dsH^{p}_T\coloneqq\dsH^{p,1}_T $ and $\dsH^{p}_{T,
  loc}\coloneqq\dsH^{p,1}_{T, loc} $.

\section{A Risk-Neutral Agent under Incremental Cash-flow}
\label{app:sec:risk.neutral}
In this section, we will derive an optimal collateral with a risk-neutral Agent
A: $U_A(x) = x$. As in previous sections, the Agent B is risk-averse as
$U_B(x) = -e^{-\gamma^Bx}$. In this case, we can relax the assumption on margin
funding rate of $A$. Then, we intend to derive similar arguments as in
\cref{sec:analysis} and \cref{sec:margin} with assuming   
\begin{align}
  s^{A,m}>0. \label{app:funding.condition}
\end{align}. Now, to model the
incremental cash-flow, assume that the bank has had contracts given by some
endowed c\`adl\`ag $\dsF$-adapted processes $(\frD^E, e^E, m^E)$ before
initiation of the new contract. If the two parties do not enter the new
contract, the cash-flow remains as
\begin{align}
 \frC^E_t = \1_{\tau > t}\frD^E_t + \1_{\tau \leq t}\big(\frD^E_{\tau} + e^E_{\tau}\big)
  -\1_{\tau =\tau^A\leq t}L^A(e^E_{\tau} - m^E_{\tau-})^+ + \1_{\tau=\tau^B\leq t}L^B(e^E_{\tau} - m^E_{\tau-})^-.\nonumber
\end{align}
On the other hand, with the new contract, the exposure and margin become
$(e^E + e )$ and $(m^E+m )$, respectively. Therefore, with the new contract,
the summed cash-flows are
\begin{align}
  \frC^S_t\coloneqq
  &  \1_{\tau > t}(\frD^E_t+\frD _t) + \1_{\tau \leq t}\big(\frD^E_{\tau}+\frD _\tau +
    e^E_{\tau}+e _\tau\big)
    \nonumber\\
  &-\1_{\tau =\tau^A\leq t}L^A(e^E_{\tau}+e _\tau -
    m^E_{\tau-}-m _{\tau-})^+ 
  + \1_{\tau=\tau^B\leq t}L^B(e^E_{\tau}+e _\tau - m^E_{\tau-}-m _{\tau-})^-. \nonumber
\end{align}
Thus, the amount that should be dealt with by the Agent A is the increment from
$\frC^E$ to $\frC^S$, namely for $t \leq T$, 
\begin{align}
  \frC_t \coloneqq
  &\frC^S_t -\frC^E_t\nonumber\\
  =&\frD _{t\wedge(\tau-)} + \1_{\tau \leq t} e _{\tau-}
    -\1_{\tau =\tau^A\leq t}L^A\big((e _{\tau-} - m _{\tau-} + e^E_{\tau} - m^E_{\tau-})^+
    -(e^E_{\tau} - m^E_{\tau-})^+\big) \nonumber\\
          &+ \1_{\tau=\tau^B\leq t}L^B\big((e _{\tau-} -
            m _{\tau-} + e^E_{\tau} - m^E_{\tau-})^- - (e^E_{\tau} -
            m^E_{\tau-})^{-}\big). \label{app:cashflow} 
\end{align} 
Thus, we denote the amount of breach of the contract as
\begin{align}
  \Theta_t(\delta) = \1_{\tau^A = t}L^A\big((\delta+\delta_t^E)^+-(\delta_t^E)^{+}\big)
    -\1_{ \tau^B = t}L^B\big((\delta+\delta_t^E)^--(\delta_t^E)^{-}\big). \label{app:breach}
\end{align}
Moreover, by \cref{app:funding.condition}, the $\dsF$-reduction of
$\Vb^A$, which is derived in \cref{theorem.reduced}, becomes slightly different
as
\begin{align}
  \df v^{A, \delta}_t =& (s^{A, m}_t\delta_t+\alpha^A_t)\df t +
                    \phi^A_t\df W_t \nonumber\\
  \alpha_t^A\coloneqq&s^{A, \Delta}_tv_t + \phi^A_t\Lambda^A_t + \Delta^A_tb^A_t \nonumber\\
  s^{A, \Delta}_t\coloneqq& s^A_t - s^{A, m}_t.\nonumber
\end{align}
Notice that $v^A$ depends on $\delta$ since we assumed that $s^{A, m}>0$, which is
the main mathematical difference from the main sections.  We still assume that
the Agent B can deliver the collateral without any excessive cost/benefit, i.e.,
$s^{B,m}=0$. In this case, $v^B$ does not depend on $\delta $.  Because of the
dependence between $v^A$ and $\delta$, we impose a slightly stronger condition for
the admissible set of collateral $\calD$:
\begin{align}
  \dsE \bigg[\big\vert G_TU_A(v^{A, \delta}_T)\big\vert
  + \big\vert\lambda G_TU_B(v^{B}_T)\big\vert +\int_{0}^{T}\big\vert g_t(v^{A, \delta}_t, v^{B}_t,
  \delta_t)\big\vert\df t\bigg]<\infty, \label{app:admissible.condition}
\end{align}
where we now denote
$g_t \coloneqq \1_{\delta +\delta^E\geq 0}g^+_t + \1_{\delta+\delta^E < 0}g^-_t$ and
\begin{align}
  g^+_t(v^A, v^B,  \delta) \coloneqq
     & G_t\big[h^A_tU_A\big (v^A + L^A(\delta-(\delta^E_t)^-)\big)
       +\lambda h^A_tU_B\big(v^B-L^AK_t(\delta-(\delta^E_t)^-)\big) \nonumber\\
     &\quad\quad+h_t^B\big(U_A(v^A+L^B(\delta^E_t)^-) + \lambda U_B(v^B-L^BK_t(\delta^E_t)^-
       )\big)\big], \nonumber
     \\
     g^-_t( v^A, v^B, \delta) \coloneqq
     & G_t\big[h_t^BU_A\big(v^A +L^B(\delta+(\delta^E_t)^+)\big)
       + \lambda h^B_tU_B\big(v^B - L^BK_t(\delta+(\delta^E_t)^+)\big)\nonumber\\
     &\quad \quad~+h_t^A\big(U_A(v^A-L^A(\delta^E_t)^+) + \lambda U_B(v^B+L^AK_t(\delta^E_t)^+)
       \big)\big].\nonumber
\end{align}
As in \cref{sec:optimal.collateral}, the first task is to characterize the
optimal collateral by MOP.  To this end, we slightly modify \cref{sold} by
merging the one terminal condition $G_Tv_T^{A, \delta }$ into $\df t$-integral
term. Observe that It\^o's formula yields
\begin{align}
  \df \big(G_tv_t^{A, \delta }\big)
  = G_t\big[s^{A, m}_t\delta _t +\alpha^A_t-h^0_tv_t^A
  \big]\df t +
  G_tv_t^{A, \delta }\phi^A_t \df W_t.
\end{align}
If $ Gv^{A, \delta }\phi^A \in \dsH^2_{T}$, the It\^o's integral term is an
$(\dsP, \dsF)$-local martingale. Thus,
\begin{align}
  \dsE\big[G_Tv_T^{A, \delta } - (\nu^A+p)\big]
  = \dsE\bigg[\int_{0}^{T}G_t\big[s^{A, m}_t\delta _t +\alpha^A_t-h^0_tv_t^A
      \big]\df t
      \bigg]. \nonumber
\end{align}
Thus, (\ref{sold}) can be written as
\begin{align}
  \max_{\delta \in \calD}\dsE\bigg[ (\nu^A+p)&+\lambda G_TU_B(v^{B}_T) \nonumber\\
  &+ \int_{0}^{T}\big[g_t(v^{A, \delta }_t, v^B_t,\delta _t)+
  G_t(s^{A, m}_t\delta _t +\alpha^A_t-h^0_tv_t^{A, \delta })
  \big]\df t\bigg].    \label{target.lin.exp.pre}
\end{align}
Then, we define a dynamic version of \cref{target.lin.exp.pre} as:
\begin{align}
  J^\varepsilon_t(p)\coloneqq\esssup_{\delta \in \calD(t, \varepsilon)}\dsE\bigg[
  (&\nu^A+p)+\lambda G_TU_B(v^{B}_T) \nonumber\\
  &+
  \int_{t}^{T}\Big[g_s(v^{A, \delta }_s, v^B_s,\delta _s)+
  G_s(s^{A, m}_s\delta _s +\alpha^A_s-h^0_sv_s^{A, \delta })
  \Big]\df s \bigg|\calF_t\bigg]. \label{line.exp.int}
\end{align}
Then  $(J^\varepsilon_t)_{0\leq t\leq T}$ is chosen as a c\`adl\`ag version such
that  for any $\varepsilon \in \calD$,
\begin{align}
  \Big\{J^\varepsilon_t + \int_{0}^{t}\big[g_s(v^{A, \varepsilon}_s, v^B_s,\varepsilon_s)+
  G_s(s^{A, m}_s\varepsilon_s +\alpha^A_s-h^0_sv_s^{A, \varepsilon})
  \big]\df s\Big\}_{0 \leq t \leq T}\nonumber 
\end{align}
is an $(\dsP, \dsF )$-supermartingale. Moreover, for the optimal collateral $\delta^*$ for
$J_0$,
\begin{align}
  \Big\{J^{\delta^*}_t + \int_{0}^{t}\big[g_s(v^{A, \delta^*}_s, v^B_s,\delta^*_s)+
  G_s(s^{A, m}_s\delta^*_s +\alpha^A_s-h^0_sv_s^{A, \delta^*})
  \big]\df s\Big\}_{0 \leq t \leq T}  \nonumber 
\end{align}
is an $( \dsP, \dsF)$-martingale.  The detail is summarized in the following
theorem. Before giving the theorem, we introduce two notations. We separate
$\delta$ from $v^A$ by denoting  $\df \vt^A_t = \df v^{A, \delta}_t - s^{A, m}_t\delta_t \df
t$, more precisely,
\begin{align}
  \vt^A_t = \int_{0}^{t}\alpha^A_s\df s+ \int_{0}^{t} \phi^A_s \df W_s.
\end{align}
Note that $\vt^A_0 = 0$. In addition, we denote 
\begin{align}
  I_t\coloneqq \int_{t}^{T} G_sh^\Delta_s\df s. 
\end{align}
The optimal collateral will later be represented by $(I_t)_{t\geq0}$. This term
appears in this section since $s^{A, m}$ can be positive.  If we consider the
cost of delivering collateral, then when to default becomes also
important. However, the effect of default time can be still marginal.  Recall
the definition that $h^\Delta = h - h^0$. Therefore, $(I_t)_{t\geq0}$ can be understood
as a correcting term of collateral for the dependence of default times. When
$\tau^A$ and $\tau^B$ are independent, we have $h = h^0$ and   $I = 0$.
\begin{theorem}\label{theorem.le}
  Assume that  $Gh^\Delta$ is deterministic. Define
  \begin{align}
    \widehat{\delta}_t(\vt^A_t, v^B_t)\coloneqq
    &\argmax_{\delta \in A}\big[\gt_t(\vt^{A}_t, v^{B}_t,   \delta)
      +I_ts^{A,m}_t\delta\big],  \label{deltastar.le} \\
      \gt_t(v^A, v^B, \delta)  \coloneqq
  & \1_{\delta + \delta^E \geq0} \gt^+_t(v^A, v^B, \delta)  + \1_{\delta + \delta^E \geq0} \gt^+_t(v^A, v^B, \delta),
    \nonumber\\
  \gt^i_t(v^A, v^B, \delta)
  \coloneqq & g^i_t(v^A, v^B, \delta) +G_t\big(s^{A, m}_t\delta+\alpha^A_t-h^0_tv^A \big),~~i
              \in \{-,+\}. \nonumber\\
 \nonumber
\end{align}
If $\csol\in \calD$, then
$(\widehat{\delta}_t(\vt^A_t, \vt^B_t))_{0\leq t \leq T}$ is a solution of (\ref{sold}).
\begin{proof}
For $\varepsilon \in \calD$, we define an $(\dsP, \dsF)$-semimartingale $(Y_t)_{t\geq0}$ as
\begin{align}
  Y_t \coloneqq& J^\varepsilon_t - \dsE\bigg[\int_{t}^{T}G_uh^\Delta_u \bigg(\int_{0}^{t}s^{A, m}_s\varepsilon_s\df s
                 \bigg)\df u \bigg|\calF_t\bigg]  \nonumber\\
  =& J^\varepsilon_t - \dsE\bigg[ \int_{0}^{t}\bigg(\int_{t}^{T}G_uh^\Delta_u\df u \bigg)
     s^{A, m}_s\varepsilon_s\df s  \bigg| \calF_t\bigg]  \nonumber\\
  =&J^\varepsilon_t - \int_{0}^{t}I_ts^{A, m}_s\varepsilon_s \df s.\nonumber 
\end{align}
Notice that $Y$ does not depend on $\varepsilon$. We also define
\begin{align}
  \xi^\varepsilon_t \coloneqq Y_t+ \int_{0}^{t}I_ts^{A, m}_s\varepsilon_s \df s
  + \int_{0}^{t}\big[g_s(v^{A, \varepsilon}_s, v^{B}_s, \varepsilon_s)+G_s(s^{A, m}_s\varepsilon_s
  +\alpha^A_s-h^0_sv_s^{A, \varepsilon})\big]\df s.  \label{lin.exp.crude1}
\end{align}
To simplify (\ref{lin.exp.crude1}), note that
\begin{align}
  g_t(v^{A, \varepsilon}_t, v^B_t,&\varepsilon_t)+G_t(s^{A, m}_t\varepsilon_t  +\alpha^A_t-h^0_tv_t^{A, \varepsilon}) \nonumber\\
  = &\big[g_t(v^{A, \varepsilon}_t, v^B_t,\varepsilon_t) - G_th_tv^{A, \varepsilon}_t\big] + G_th_tv^{A, \varepsilon}_t-h^0_tv_t^{A, \varepsilon}
      +G_t(s^{A, m}_t\varepsilon_t  +\alpha^A_t)
      \nonumber\\
  = &\big[g_t(v^{A, \varepsilon}_t, v^B_t,\varepsilon_t) - G_th_tv^{A, \varepsilon}_t\big] + G_th^\Delta_tv^{A, \varepsilon}_t
      +G_t(s^{A, m}_t\varepsilon_t  +\alpha^A_t).
      \nonumber
\end{align}
In addition, by Fubini's theorem,
\begin{align}
  \int_{0}^{t}G_uh^\Delta_u\Big[\int_{0}^{u}s^{A,m}_s\varepsilon_s\df s\Big]\df u =
   \int_{0}^{t}s^{A,m}_s\varepsilon_s\Big[\int_{s}^{t}G_uh^\Delta_u\df u\Big]\df s.\nonumber
\end{align}
Moreover, 
\begin{align}
  \int_{0}^{t}G_uh^\Delta_uv^{A, \varepsilon}_u \df u
  =& \int_{0}^{t}G_uh^\Delta_u\Big[v^{A, \varepsilon}_u-\int_{0}^{u}s^{A, m}_s\varepsilon_s\df s\Big]\df u
     + \int_{0}^{t}G_uh^\Delta_u\Big[\int_{0}^{u}s^{A, m}_s\varepsilon_s\df s\Big] \df u \nonumber\\
   =& \int_{0}^{t}G_uh^\Delta_u\vt^{A}_u\df u
  + \int_{0}^{t}s^{A,m}_s\varepsilon_s\Big[\int_{s}^{t}G_uh^\Delta_u\df u\Big]\df s,
\end{align}
and for $t \leq T$, 
\begin{align}
  g_t(v^{A, \varepsilon}_t, v^B_t,\varepsilon_t) - G_th_tv_t^{A, \varepsilon} +  Gh_t\vt_t^{A}
  = g(\vt^A_t, v^B_t, \varepsilon_t).
\end{align}
Therefore, (\ref{lin.exp.crude1}) can be rewritten as
\begin{align}
 Y_t+ \int_{0}^{t}I_ts^{A, m}_s&\varepsilon_s \df s
  + \int_{0}^{t}\big[g_s(v^{A, \varepsilon}_s, v^{B}_s, \varepsilon_s)+G_s(s^{A, m}_s\varepsilon_s
  +\alpha^A_s-h^0_sv_s^{A, \varepsilon})\big]\df s\nonumber\\
  =& Y_t+ \int_{0}^{t}I_ts^{A, m}_s\varepsilon_s \df s
     +\int_{0}^{t}s^{A,m}_s\varepsilon_s\Big[\int_{s}^{t}G_uh^\Delta_u\df u\Big]\df s\nonumber\\
    & + \int_{0}^{t} \Big(g_s(v^{A, \varepsilon}_s, v^B_s,\varepsilon_s) - G_sh_sv_s^{A, \varepsilon} +
      Gh^\Delta_sv_s^{A} +G_s(s^{A, m}_s\varepsilon_s  +\alpha^A_s)\df s\nonumber\\
  =&Y_t+\int_{0}^{t}I_ss^{A, m}_s\varepsilon_s \df s
     +\int_{0}^{t} \Big(g_s(\vt^{A}_s, v^B_s,\varepsilon_s)  +G_s(s^{A, m}_s\varepsilon_s
     +\alpha^A_s-h^0\vt^A_s) \Big)\df s
     \nonumber\\
  =&Y_t+\int_{0}^{t}I_ss^{A, m}_s\varepsilon_s \df s
     +\int_{0}^{t} \gt_s(\vt^{A}_s, v^B_s,\varepsilon_s)\df s.\nonumber
\end{align}
Then, since $Y$ is independent of $\varepsilon\in \calD$, by the assumption of
admissibility of $\csol$, for any $\varepsilon\in \calD$,
 we have that $ \xi^\varepsilon - \xi^{\csol}$ 
is an $(\dsP, \dsF)$-supermartingale. It follows that for any $\varepsilon \in
\calD$, 
$\dsE\big[  \xi^\varepsilon_T - \xi^{\csol}_T\big] \leq \dsE\big[\xi^\varepsilon_0 - \xi^{\csol}_0\big]=0$. 
Then, by the admissibility, $\csol$ is a solution of \cref{sold}.
\end{proof}
\end{theorem}
The last step is to show $\csol$ is admissible given some conditions. We
consider $A = \dsR$ and find the explicit form of $\csol$ for the case. Then, the
integrability condition is easy to check. First, notice that $\gt^+$, $\gt^-$
are continuously differentiable and strictly concave in $\delta$. Thus, for any
$(t, v^A, v^B)$, there exists $\widehat{I}^i_t(v^A, v^B)$, $i \in \{-, +\}$ such that
\begin{align}
  \partial_\delta\gt^i_t(v^A, v^B, \widehat{I}^i_t(v^A, v^B)) + s^{A, m}_tI_t=0
\end{align}
Then, it is easy to check that $\csol$ is attained at $I^-$, $I^+$, and
$-\delta^E$.  Observe the
precise forms of $\gt^i$, $i \in \{-, +\}$, are
\begin{align}
  \gt^-_t(v^A, v^B, \delta)\coloneqq
            &G_t\big[h^\Delta_tv^A  + (h^B_tL^B + s^{A, m}_t)\delta+ \alpha^A_t
              +(h^B_tL^B - h^A_tL^A)(\delta^E_t)^+
            \nonumber\\  
            &+ \lambda h^B_tU_B\big(v^B - L^BK_t(\delta+(\delta^E_t)^+)\big)
              + \lambda U_B(v^B+L^AK_t(\delta^E_t)^+)
              \big)\big],\nonumber\\
  \gt^+_t(v^A, v^B, \delta)
  \coloneqq &G_t\big[h^\Delta_tv^A  + (h^A_tL^A + s^{A, m}_t)\delta+ \alpha^A_t
              +(h^B_tL^B - h^A_tL^A)(\delta^E_t)^- \nonumber\\
       &+\lambda h^A_tU_B\big(v^B-L^AK_t(\delta-(\delta^E_t)^-)\big) 
         + \lambda h^B_tU_B(v^B-L^BK_t(\delta^E_t)^-
       )\big)\big]. \nonumber  
\end{align}
Therefore, assuming $h^iL^i >0$, $i\in \{A, B\}$, $I^i$, $i \in \{-, +\}$, can be explicitly represented as
\begin{align}
   \widehat{I}^-_t(v^A, v^B)=& -(\delta^E_t)^++ \frac{ v^B}{K_tL^B}
  +\frac{1}{\gamma^BK_tL^B}\ln{\bigg(\frac{ G_t[h^B_tL^B + s^{A, m}_t] + s^{A, m}_tI_t}
  {G_t\lambda\gamma^BK_t h^B_tL^B}\bigg)}, \label{def.im.le}\\
 \widehat{I}^+_t(v^A, v^B)=& (\delta^E_t)^-+ \frac{ v^B}{K_tL^A}
  +\frac{1}{\gamma^BK_tL^A}\ln{\bigg(\frac{ G_t[h^A_tL^A + s^{A, m}_t] + s^{A,
                   m}_tI_t} 
  {G_t\lambda\gamma^BK_t h^A_tL^A}\bigg)}.  \label{def.ip.le}
\end{align}
Then \cref{theorem.le} is completed by the next lemma. The proof is similar to
that of \cref{lemma.ee}, so we omit it.
\begin{lemma}\label{lemma.le}
Let $A = \dsR$. Assume that $\delta^E, e, Z, \pi^i, ~i \in \{A, B\}$, are
bounded. Moreover, assume $h^iL^i \in\dsH^2_T$ and 
\begin{align}
  \frac{(G+I)s^{A,m}}{Gh^iL^i}, ~i \in \{A, B\}, \label{le.bound}
\end{align}
are bounded. Then $\csol \in \calD$.
\end{lemma}
Different from \cref{def.im.ee} and \cref{def.ip.ee}, $\csol$ depends on $h^A$
and $h^B$. As mentioned, this dependence arises from the funding impact of
$s^{A, m}$, and by setting $s^{A, m} = 0$, \cref{def.im.le}-\cref{def.im.le} are
reduced to
\begin{align}
   \widehat{I}^-_t(v^A, v^B)=& -(\delta^E_t)^++ \frac{ v^B}{K_tL^B}
  -\frac{\ln{(\lambda\gamma^BK_t)}}{\gamma^BK_tL^B}, \label{def.im.le.red}\\ 
 \widehat{I}^+_t(v^A, v^B)=& (\delta^E_t)^-+ \frac{ v^B}{K_tL^A}
  -\frac{\ln{(\lambda\gamma^BK_t)}}{\gamma^BK_tL^A}.  \label{def.ip.le.red}
\end{align}
In addition, we can derive similar interpretations as in \cref{sec:analysis}.
In what follows, we moreover, assume that all parameter are
constant and the default times are independent on $\dsF$, i.e., $I=0$.
The full collateral convention can be said that $\delta^* =v  - c^*= 0 $ and $\delta^E = 0$,
$\df \dsP \otimes \df t$-a.s. Therefore, $I^i = 0$, for any $t\leq T$.
By (\ref{def.im.le}) and (\ref{def.ip.le}), full collateralization
requires that
\begin{align}
  v^B_t - \frac{1}{\gamma^B}(s^A-s^B)t = \frac{1}{\gamma^B}\ln{\Big(\frac{\lambda \gamma^B
  h^iL^i}{h^iL^i+s^{A, m}}\Big)}, ~~i \in \{A, B\}, ~~\dpdt.   \label{vcconstant}
\end{align}
In particular, $(v^B_t - (s^A-s^B)t/\gamma^B)_{t\geq0}$ should be constant. Thus,
(\ref{vcconstant}) implies that $\phi^B =\pi^B + \Delta^B= 0$, i.e., delta-hedge, and
\begin{align}
  -s^{B}Kv  +  \phi^B\Lambda^B_t -\Delta^Bb^B- \frac{s^A-s^B}{\gamma^B}  =0, ~\dpdt.
  \label{le.drift.zero}
\end{align}
Consider a contract such that $Z  \not= 0$, so necessarily $v  \not=0$ and
$\Delta^B\not=0$. Since (\ref{le.drift.zero}) should hold for all contracts such
that $Z  \not= 0$, (\ref{le.drift.zero}) implies that
 $ s^{B}=  b^B =  s^A-s^B =0 $.
Equivalently, by (\ref{def.fs}), (\ref{bs}), 
  $s^B = s^A = s^{A, m}=0$. 
Therefore, the margin requirement hinges on the assumption of absence of funding
impacts and delta-hedge of the Agent B. No property of the Agent A's hedging
strategy was derived since we assumed the Agent A is risk-neutral. 
\section{Proofs of Lemmas}
\label{app:sec:proof}
\begin{proof} [Proof of \cref{lem:em}]
(\rom{1}) is from the definition and (\rom{2}) is a directly obtained from
(\rom{1}).  For (\rom{3}), notice that
$B_t^{-1}e _t + \int_{0}^{t}B^{-1}_s\df \frD _s$ is an $(\dsQ, \dsF)$-local
martingale. Thus, by (local) martingale representation property,
there exists $Z  \in \dsH^{2, d}_{T, loc}$ such that for any $t\geq0$, 
\begin{align}
B_t^{-1}e _t + \int_{0}^{t}B^{-1}_s\df \frD _s = \int_{0}^{t}Z _s\df
  W^{\dsQ}_s,\nonumber 
\end{align}
where $W^{\dsQ}$ is the Brownian motion under $\dsQ$, i.e., 
$ W^{\dsQ}_t = W_t + \int_{0}^{t}\Lambda_s\df s$.
Therefore, $(e_t)_{t\geq0} $ follows the SDE:
\begin{align}
  \df e _t =& r_te _t\df t + B_tZ _t\df W^{\dsQ}_t - \df \frD _t \nonumber\\
  =& \big(r_te _t +  B_tZ _t\Lambda_t\big)\df t  +B_tZ _t\df W_t - 
  \df \frD _t. \label{dyn:e}
\end{align}
By (\rom{3}), $(e_t)_{t\geq0} $ is an $\dsF$-adapted c\`adl\`ag process, but $\tau$ avoids
$\dsF$-stopping times. Thus, $\Delta e _{\tau}=0$ almost surely, equivalently $e _{\tau-}
= e _{\tau}$ a.s.        
\end{proof}
\begin{proof} [Proof of \cref{lemma.ee}]
  Let $i \in \{A, B\}$ and $\Psi(x) :=e^{ C x }$ for $C \in \dsR$. It suffices to show
  that for any $C \in \dsR$, $\Psi(X)$, $\Psi(v^i)$, are in $\dsS^2_T$. Note that
  $v = (B^A)^{-1}e$ is bounded and, by \cref{delta} and \cref{vf}, $\Delta^i$ and
  $\phi^i$ are also bounded. Denoting 
 \begin{align}
   \alpha^A \coloneqq& \phi^A\Lambda^A + \Delta^Ab^A + s^Av\nonumber\\
   \alpha^B \coloneqq& \phi^B\Lambda^B- \Delta^Bb^B - Ks^Bv,\nonumber
 \end{align}
we can write $\df v^i_t = \alpha^i_t\df t+\phi^i_t \df W_t$. Applying It\^o's formula to
$\Psi(v^i)$,
\begin{align}
  \df \Psi(v^i_t) = \big(C\alpha^i_t + (C\phi^i_t)^2/2\big)\Psi(v^i_t)\df t +C\phi^i_t\Psi(v^i_t)
  \df W_t.
  \label{lem.ee.sde}
\end{align}
By the assumptions, the coefficients in \cref{lem.ee.sde} are uniformly Lipsitch 
continuous. Thus, there exists a unique solution of \cref{lem.ee.sde} such that 
\begin{align}
 \dsE\Big[\sup _{t \leq T}|\Psi(v^i_t)|^2\Big] < \infty.   \nonumber 
\end{align}
In particular, $U_i(v^i)\in \dsS^2_T$, so we obtain the integrability condition
\cref{condition.integrability}.
It is similarly obtained that  $\Psi(X)\in \dsS^2_T$.
Let $(C_t)_{t\geq0}$ be an arbitrary bounded deterministic process. Then, we also have
$\exp{(C X)} \in \dsS^2_T$ and it follow that
\begin{align}
  &U_A\bigg(\frac{-\gamma^A}{\gamma^BK + \gamma^A}X\bigg) \in \dsS^2_T \nonumber\\
  &U_B\bigg(\frac{K\gamma^A}{\gamma^BK + \gamma^A}X\bigg) \in \dsS^2_T. \nonumber
\end{align}
Thus, by \cref{sol.ee}-\cref{psic}, \cref{def.im.ee}, and \cref{def.ip.ee}, we
obtain $\delta^*(p, X) \in \calD$. 
\end{proof}
\begin{proof}[Proof of \cref{lem:col}]
Recall from \cref{def:f-} and \cref{def:f+} that
\begin{align}
  \partial_\delta f^+(t, p, x, \delta)   \coloneqq
  &  h_t^AL^A\big[-\gamma^AU_A(x +L^A\delta) + \lambda \gamma^BK_tU_B(\nu^B-p-L^AK_t\delta)\big],\nonumber \\
  \partial_\delta f^-(t, p, x, \delta)   \coloneqq
  &  h_t^BL^B\hspace{0.065cm}\big[-\gamma^AU_A(x+L^B\delta) + \lambda
    \gamma^BK_tU_B(\nu^B-p-L^BK_t\delta)\big]. \nonumber 
\end{align}
Let $I^i_t(p, x)$ denote the function such that $\partial_\delta f^i(t, p, x, I^i_t(p, x))
=0$. Since $f^i$, $i \in \{-, +\}$, are concave w.r.t $\delta$, $I^i$ uniquely
exists. Let us denote
\begin{align}
  \ft^+(t,p, x) \coloneqq& \max_{0\leq \delta }f^+(t, p, x, \delta),\nonumber\\
  \ft^-(t,p, x) \coloneqq& \max_{ \delta\leq 0 }f^-(t, p, x, \delta),\nonumber\\
  \ft(t,p,x ) \coloneqq& \max_{ \delta\in \dsR }f(t, p, x, \delta).\nonumber 
\end{align}
Then it follows that
\begin{align}
  \ft(t, p, x) \coloneqq&f(t, p, x, \delta^*(t, p, x)),\nonumber\\
  =&\ft^+(t, p, x)\1_{\ft^+(t, p, x)\geq \ft^-(t, p, x)} +\ft^-(t, p, x)
     \1_{\ft^+(t, p, x)\leq \ft^-(t, p, x)}. \nonumber  
\end{align}
Thus, for finding $\delta^*$, we should characterize the region that
$\ft^+(t, p, x)\geq \ft^-(t, p, x)$. The solutions $\delta^i$, $i \in \{-,+\}$, of $\ft^i$
can be found explicitly:
\begin{align}
  \delta^+_t( p, x) &= \left\lbrace
  \begin{array}{r@{}l}
    0,~~~~~~~~&  0 > \partial_\delta f^+(t, p, x, 0),\\
    I^+_t( p, x),~~~~~~~~&0 \leq\partial_\delta f^+(t, p, x, 0), 
  \end{array} \nonumber 
  \right.\\
  \delta^-_t( p, x) &= \left\lbrace
    \begin{array}{r@{}l}
 I^-_t( p, x),~~~~~~~~&\partial_\delta f^-(t, p, x, 0) \leq 0,   \\
    0,~~~~~~~~&  0 < \partial_\delta f^-(t, p, x, 0).
  \end{array}
                \right. \nonumber 
\end{align}
Moreover,  notice that since $h^i \geq 0$ and $L^i \geq 0$,
  $\partial_\delta f^+(0) * \partial_{\delta}f^-(0) \geq 0$.
The rest of the proof is merely a straightforward comparison of $\ft^i$ in each
region. In what follows, we suppress $t, x, p$.
\begin{enumerate}[label=(\Roman*)] 
\item Let $ 0 \leq \partial_\delta f^+(0)\wedge \partial_\delta f^-(0)$. In other words, 
\begin{align}
\gamma_Ax + \gamma_Bp \leq \gamma_B\nu^B - \ln{\big(\lambda K_t\gratio\big)}.\nonumber 
\end{align}
Thus, $I^i \geq 0$, $i \in \{-, +\}$, and $\delta^- =0$.  Moreover, 
\begin{align}
 \ft^- - \ft^+ = f^-(0) - f^+(I^+) =f^+(0) - f^+(I^+ )\leq0. \nonumber 
\end{align}
Hence, $\delta^* =\delta^+= I^+  \geq0$. 
\item Let $ 0 > \partial f^+(0) \vee \partial f^-(0)$. Then, $\delta^+ = 0$ and $I^- \leq0$.  Therefore, by
  similar calculation, $\delta^* =\delta^-= I^- \leq 0$.
\end{enumerate}
\end{proof}
\end{appendices}

\section*{Acknowledgments}
We thank St\'ephane Cr\'epey for
    spending much time to help us improve this paper.

\bibliographystyle{siamplain}
\bibliography{arxiv.rs}   

\begin{thebibliography}{10}

\bibitem{agarwal2019numerical}
{\sc A.~Agarwal, S.~De~Marco, E.~Gobet, J.-G. Lopez-Salas, F.~Noubiagain, and
  A.~Zhou}, {\em Numerical approximations of mckean anticipative backward
  stochastic differential equations arising in initial margin requirements},
  ESAIM: Proceedings and Surveys, 65 (2019), pp.~1--26.

\bibitem{albanese2014accounting}
{\sc C.~Albanese and L.~Andersen}, {\em Accounting for {OTC} derivatives:
  Funding adjustments and the re-hypothecation option}, Available at
  SSRN:2482955,  (2014).

\bibitem{albanese2013restructuring}
{\sc C.~Albanese, D.~Brigo, and F.~Oertel}, {\em Restructuring counterparty
  credit risk}, International Journal of Theoretical and Applied Finance, 16
  (2013), p.~1350010.

\bibitem{albanese2018wealth}
{\sc C.~Albanese, M.~Chataigner, and S.~Cr{\'e}pey}, {\em Wealth transfers,
  indifference pricing, and {XVA} compression schemes}, Available at
  https://math.maths.univ-evry.fr/crepey/papers/wealth\%20transfer-NEW.pdf,
  (2018).

\bibitem{andersen2019funding}
{\sc L.~Andersen, D.~Duffie, and Y.~Song}, {\em Funding value adjustments}, The
  Journal of Finance, 74 (2019), pp.~145--192.

\bibitem{bichuch2016arbitrage1}
{\sc M.~Bichuch, A.~Capponi, and S.~Sturm}, {\em Arbitrage-free pricing of
  {XVA}-part {I}: Framework and explicit examples}, Available at
  arXiv:1501.05893,  (2016).

\bibitem{bichuch2016arbitrage2}
{\sc M.~Bichuch, A.~Capponi, and S.~Sturm}, {\em Arbitrage-free pricing of
  {XVA}-part {II}: {PDE} representation and numerical analysis}, Available at
  arXiv:1502.06106,  (2016).

\bibitem{bichuch2017arbitrage}
{\sc M.~Bichuch, A.~Capponi, and S.~Sturm}, {\em Arbitrage-free {XVA}},
  Mathematical Finance, 28 (2018), pp.~582--620.

\bibitem{bichuch2018robust}
{\sc M.~Bichuch, A.~Capponi, and S.~Sturm}, {\em Robust {XVA}}, Available at
  arXiv:1808.04908,  (2018).

\bibitem{bielecki2018arbitrage}
{\sc T.~R. Bielecki, I.~Cialenco, and M.~Rutkowski}, {\em Arbitrage-free
  pricing of derivatives in nonlinear market models}, Probability, Uncertainty
  and Quantitative Risk, 3 (2018), pp.~1--56.

\bibitem{bielecki2008pricing}
{\sc T.~R. Bielecki, M.~Jeanblanc, and M.~Rutkowski}, {\em Pricing and trading
  credit default swaps in a hazard process model}, The Annals of Applied
  Probability, 18 (2008), pp.~2495--2529.

\bibitem{bielecki2015valuation}
{\sc T.~R. Bielecki and M.~Rutkowski}, {\em Valuation and hedging of contracts
  with funding costs and collateralization}, SIAM Journal on Financial
  Mathematics, 6 (2015), pp.~594--655.

\bibitem{bis2015margin}
{\sc BIS and IOSCO}, {\em Margin requirements for non-centrally cleared
  derivatives}, 2015.

\bibitem{bo2017portfolio}
{\sc L.~Bo}, {\em Portfolio optimization of credit swap under funding costs},
  Probability, Uncertainty and Quantitative Risk, 2 (2017), p.~12.

\bibitem{bo2016optimal}
{\sc L.~Bo and A.~Capponi}, {\em Optimal credit investment with borrowing
  costs}, Mathematics of Operations Research, 42 (2016), pp.~546--575.

\bibitem{bo2017credit}
{\sc L.~Bo, A.~Capponi, and P.~Chen}, {\em Credit portfolio selection with
  decaying contagion intensities}, Mathematical Finance,  (2018), pp.~1--37.

\bibitem{bo2019locally}
{\sc L.~Bo and C.~Ceci}, {\em Locally risk-minimizing hedging of counterparty
  risk for portfolio of credit derivatives}, Applied Mathematics \&
  Optimization,  (2019), pp.~1--52.

\bibitem{brigo2014arbitrage}
{\sc D.~Brigo, A.~Capponi, and A.~Pallavicini}, {\em Arbitrage-free bilateral
  counterparty risk valuation under collateralization and application to credit
  default swaps}, Mathematical Finance, 24 (2014), pp.~125--146.

\bibitem{brigo2011collateral}
{\sc D.~Brigo, A.~Capponi, A.~Pallavicini, and V.~Papatheodorou}, {\em
  Collateral margining in arbitrage-free counterparty valuation adjustment
  including re-hypotecation and netting}, Avaialble at SSRN:1744101,  (2011).

\bibitem{brigo2014nonlinear}
{\sc D.~Brigo, Q.~Liu, A.~Pallavicini, and D.~Sloth}, {\em Nonlinear valuation
  under collateral, credit risk and funding costs: a numerical case study
  extending {B}lack-{S}choles}, Handbook in Fixed-Income Securities, Wiley,
  (2014).

\bibitem{brigo2011close}
{\sc D.~Brigo and M.~Morini}, {\em Close-out convention tensions}, Risk,
  (2011), pp.~74--78.

\bibitem{brigo2014ccp}
{\sc D.~Brigo and A.~Pallavicini}, {\em Nonlinear consistent valuation of ccp
  cleared or csa bilateral trades with initial margins under credit, funding
  and wrong-way risks}, Journal of Financial Engineering, 1 (2014), p.~1450001.

\bibitem{burgard2010partial}
{\sc C.~Burgard and M.~Kjaer}, {\em Partial differential equation
  representations of derivatives with bilateral counterparty risk and funding
  costs}, Journal of Credit Risk, 7 (2010).

\bibitem{carassus2009portfolio}
{\sc L.~Carassus and H.~Pham}, {\em Portfolio optimization for piecewise
  concave criteria functions}, The 8th Workshop on Stochastic Numerics,
  (2009).

\bibitem{castagna2012yes}
{\sc A.~Castagna}, {\em Yes, {FVA} is a cost for derivatives desks}, Available
  at SSRN:2141663,  (2012).

\bibitem{chen2018forward}
{\sc J.~Chen, J.~Ma, and H.~Yin}, {\em Forward-backward {SDE}s with
  discontinuous coefficients}, Stochastic Analysis and Applications, 36 (2018),
  pp.~274--294.

\bibitem{coculescu2012hazard}
{\sc D.~Coculescu and A.~Nikeghbali}, {\em Hazard processes and martingale
  hazard processes}, Mathematical Finance, 22 (2012), pp.~519--537.

\bibitem{crepey2014counterparty}
{\sc S.~Cr{\'e}pey, T.~R. Bielecki, and D.~Brigo}, {\em Counterparty risk and
  funding: A tale of two puzzles}, CRC Press, 2014.

\bibitem{crepey2017capital}
{\sc S.~Cr{\'e}pey, R.~{\'E}lie, W.~Sabbagh, and S.~Song}, {\em When capital is
  a funding source: The {XVA} anticipated {BSDE}s}, tech. report, Working paper
  available at https://math. maths. univevry. fr/crepey, 2017.

\bibitem{cvitanic2018dynamic}
{\sc J.~Cvitani{\'c}, D.~Possama{\"\i}, and N.~Touzi}, {\em Dynamic programming
  apparoach to princial--agent roblems}, Finance and Stochastics, 22 (2018),
  pp.~1--37.

\bibitem{cvit2013contract}
{\sc J.~Cvitanic and J.~Zhang}, {\em Contract theory in continuous-time
  models}, Springer-{V}erlag {B}erlin {H}eidelberg, 2013.

\bibitem{el1997backward}
{\sc N.~El~Karoui, S.~Peng, and M.~C. Quenez}, {\em Backward stochastic
  differential equations in finance}, Mathematical Finance, 7 (1997),
  pp.~1--71.

\bibitem{elliott2000models}
{\sc R.~J. Elliott, M.~Jeanblanc, and M.~Yor}, {\em On models of default risk},
  Mathematical Finance, 10 (2000), pp.~179--195.

\bibitem{glasserman2018bounding}
{\sc P.~Glasserman and L.~Yang}, {\em Bounding wrong-way risk in {CVA}
  calculation}, Mathematical Finance, 28 (2018), pp.~268--305.

\bibitem{he1992semimartingale}
{\sc S.~W. He and J.~A. Yan}, {\em Semimartingale theory and stochastic
  calculus}, Taylor \& Francis, 1992.

\bibitem{hull2012fva}
{\sc J.~Hull and A.~White}, {\em The {FVA} debate}, Risk,  (2012), pp.~83--85.

\bibitem{hull2012fva2}
{\sc J.~Hull and A.~White}, {\em The {FVA} debate continues}, Risk,  (2012),
  p.~52.

\bibitem{jiao2013optimal}
{\sc Y.~Jiao, I.~Kharroubi, and H.~Pham}, {\em Optimal investment under
  multiple defaults risk: a {BSDE}-decomposition approach}, The Annals of
  Applied Probability, 23 (2013), pp.~455--491.

\bibitem{jiao2011optimal}
{\sc Y.~Jiao and H.~Pham}, {\em Optimal investment with counterparty risk: a
  default-density model approach}, Finance and Stochastics, 15 (2011),
  pp.~725--753.

\bibitem{nie2018american}
{\sc E.~Kim, T.~Nie, and M.~Rutkowski}, {\em Arbitrage-free pricing of american
  options in nonlinear markets}, Available at arXiv:1804.10753,  (2018).

\bibitem{nie2018game}
{\sc E.~Kim, T.~Nie, and M.~Rutkowski}, {\em Arbitrage-free pricing of game
  options in nonlinear markets}, Available at arXiv:1807.05448,  (2018).

\bibitem{lee2017binary}
{\sc J.~Lee and C.~Zhou}, {\em A binary nature of funding impacts in bilateral
  contracts}, Avaliable at arXiv:1703.00259,  (2017).

\bibitem{li2016fva}
{\sc C.~Li and L.~Wu}, {\em {FVA} and {CVA} for collateralized trades with
  re-hypothecation}, Wilmott, 2016 (2016), pp.~50--59.

\bibitem{modigliani1958cost}
{\sc F.~Modigliani and M.~H. Miller}, {\em The cost of capital, corporation
  finance and the theory of investment}, The American economic review, 48
  (1958), pp.~261--297.

\bibitem{murphy2013otc}
{\sc D.~Murphy}, {\em {OTC} Derivatives: Bilateral Trading and Central
  Clearing: An Introduction to Regulatory Policy, Market Impact and Systemic
  Risk}, Palgrave {M}acmillan, 2013.

\bibitem{nie2016bsde}
{\sc T.~Nie and M.~Rutkowski}, {\em A {BSDE} approach to fair bilateral pricing
  under endogenous collateralization}, Finance and Stochastics, 20 (2016),
  pp.~855--900.

\bibitem{piterbarg2010funding}
{\sc V.~Piterbarg}, {\em Funding beyond discounting: collateral agreements and
  derivatives pricing}, Risk, 23 (2010), p.~97.

\bibitem{sircar2010utility}
{\sc R.~Sircar and T.~Zariphopoulou}, {\em Utility valuation of multi-name
  credit derivatives and application to {CDO}s}, Quantitative Finance, 10
  (2010), pp.~195--208.

\bibitem{stiglitz1969re}
{\sc J.~E. Stiglitz}, {\em A re-examination of the {M}odigliani-{M}iller
  theorem}, The American Economic Review, 59 (1969), pp.~784--793.

\bibitem{wu2015cva}
{\sc L.~Wu}, {\em {CVA} and {FVA} to derivatives trades collateralized by
  cash}, International Journal of Theoretical and Applied Finance, 18 (2015),
  p.~1550035.

\bibitem{yang2017constrained}
{\sc Z.~Yang, G.~Liang, and C.~Zhou}, {\em Constrained portfolio-consumption
  strategies with uncertain parameters and borrowing costs}, Mathematics and
  Financial Economics,  (2017), pp.~1--35.

\end{thebibliography}

\end{document}